\documentclass[letterpaper,twocolumn,10pt]{article}
\usepackage{hyperref}
\usepackage{usenix}
\usepackage{graphicx}
\usepackage{multirow}
\usepackage{caption}
\usepackage{subcaption}
\usepackage{float}
\usepackage{amsmath,amsfonts}
\usepackage{xcolor}
\usepackage{xspace}
\usepackage{pifont}
\usepackage{tikz}
\usepackage{tcolorbox}

\newcommand{\attack}{$DoubleStar$\xspace}

\newcommand{\rev}[1]{{\color{black} #1}}
\newcommand{\minor}[1]{{\color{black} #1}}

\begin{document}
\date{}

\title{\Large \bf \attack: Long-Range Attack Towards Depth Estimation based Obstacle Avoidance in Autonomous Systems}


\author{
{\rm Ce Zhou}\\
Michigan State University
\and
{\rm Qiben Yan}\thanks{Corresponding author: Dr. Qiben Yan (\href{mailto:qyan@msu.edu}{\nolinkurl{{qyan@msu.edu}}})}\\
Michigan State University
 \and
 {\rm Yan Shi}\\
Michigan State University
 \and
 {\rm Lichao Sun}\\
Lehigh University
} 


\maketitle

\begin{abstract}

Depth estimation-based obstacle avoidance has been widely adopted by autonomous systems (drones and vehicles) for safety purpose. It normally relies on a stereo camera to automatically detect obstacles and make flying/driving decisions, e.g., stopping several meters ahead of the obstacle in the path or moving away from the detected obstacle. 
In this paper, we explore new security risks associated with the stereo vision-based depth estimation algorithms used for obstacle avoidance. By exploiting the weaknesses of the stereo matching in depth estimation algorithms and the lens flare effect in optical imaging, we propose \attack, a long-range attack that injects fake obstacle depth by projecting pure light from two complementary light sources. 

\attack includes two distinctive attack formats: beams attack and orbs attack, which leverage projected light beams and lens flare orbs respectively to cause false depth perception. We successfully attack two commercial stereo cameras designed for autonomous systems (ZED and Intel RealSense). The visualization of fake depth perceived by the stereo cameras illustrates the false stereo matching induced by \attack. We further use Ardupilot to simulate the attack and demonstrate its impact on drones. To validate the attack on real systems, we perform a real-world attack towards a commercial drone equipped with state-of-the-art obstacle avoidance algorithms. Our attack can continuously 
bring a flying drone to a sudden stop or drift it away
across a long distance under various lighting conditions, even bypassing sensor fusion mechanisms. Specifically, our experimental results show that \attack creates fake depth up to 15 meters in distance at night and up to 8 meters  during the daytime. To mitigate this newly discovered threat, we provide discussions 
on potential countermeasures 
to defend against \attack.

\end{abstract}

\section{Introduction}

\begin{figure}[t]
\centering
	\includegraphics[width=0.45\textwidth]{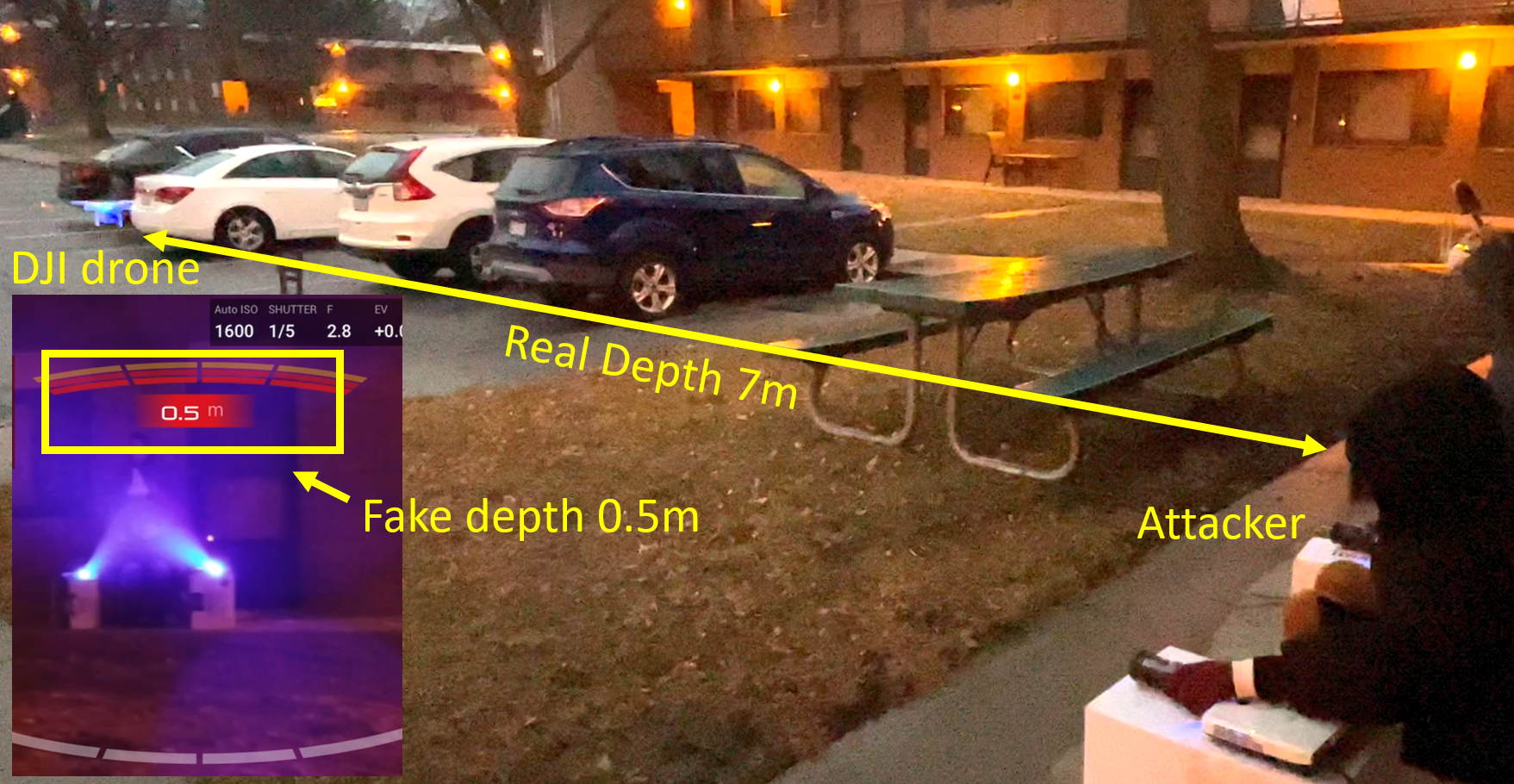}
	\caption{An attacker uses two projectors to launch \attack at 7$m$ away on a flying DJI drone. A fake depth of 0.5$m$ is created by the attack and detected by the DJI drone as a real obstacle during the daytime.}
	\label{pic1}
\end{figure}

Obstacle detection and avoidance are widely adopted in autonomous systems, such as autonomous driving vehicles~\cite{Uber, Waymo, Autopilot}, robotic vehicles~\cite{MITracecar, f1tenth}, and drones \cite{DJI_Phantom_4, Skydio_2}. Generally, Obstacle Avoidance (OA) system detects the obstacles in the surroundings via different sensors, e.g., cameras, radars, LiDARs, and ultrasonic sensors, and converts the perceived data into obstacle information (e.g., obstacle distance, obstacle type). 
The autonomous systems then make an appropriate driving/flying decision, such as raising the alarm, braking in front of the obstacle, or moving away from it. 

The recent rise in the popularity of drones and self-driving vehicles helps drive OA's prevalence, while the potential risks of OA algorithms warrants further research. Although the community produced a wealth of security research on autonomous systems over the years~\cite{son2015rocking, yan2016can, cao2019adversarial, sun2020towards, sato2020hold, shen2020drift, man2020ghostimage, nassi2020phantom}, one sensing modality that is nearly omnipresent in modern OA, the stereo camera \cite{Stereo_camera} (a.k.a., 3D depth camera), has mostly been overlooked. 
In this work, we expose the security risk of stereo cameras for the first time and propose a new attack, \emph{\attack}, which targets the depth estimation --- one of the core functionalities of stereo cameras. \attack allows an attacker to launch long-range and continuous attacks towards depth estimation by creating fake obstacles optically. Since the estimated depth is an essential input parameter to the OA systems, \attack has profound implications towards the functional safety of autonomous systems. 

\minor{
\attack builds upon a rich body of research on camera sensor security. Earlier studies that feature denial-of-service (DoS) attacks \cite{petit2015remote, truong2005preventing, yan2016can} can be detected easily by tamper detection mechanisms \cite{ribnick2006real}. \attack draws inspiration from other recent studies that have overcome such a limitation. One such attack, GhostImage \cite{man2020ghostimage}, utilizes lens flare effect to deceive the image classification systems in autonomous vehicles into misperceiving actual objects or perceiving non-existent objects. 
However, GhostImage is limited by 
the inability to sustain a continuous attack due to the pixel-level position aiming issue arisen from the white-box attack design. Our experiments further show that GhostImage is more challenging to realize against cameras with
commonly used anti-reflection coatings~\cite{afc_wiki}. 
More importantly, all existing attacks target monocular cameras.
To date, \attack is \emph{the first} to exploit stereo cameras’ vulnerabilities on autonomous systems. 
}

\minor{Stereo cameras are widely available on robotic vehicles\footnote{\minor{Robotic vehicle refers to the unmanned robot vehicle used in manufacturing, transport, military, etc.}}, which have been used for navigation and 3D tracking applications \cite{navigation , f1tenth, MITracecar}.}
Drones, on the other hand, are smaller in scale with less functional demands, whose navigation usually requires the depth information. Since LiDARs and Radars are not favorable on drones due to their form factors and high costs, the stereo camera becomes the de-facto sensor to perceive accurate depth information. 
Almost all the high-end drones are equipped with stereo cameras, such as DJI Phantom series \cite{DJI_Phantom_4}, DJI Mavic series \cite{DJI_Mavic_Series}, Skydio R1 \cite{Skydio_R1}, Yuneec Typhoon series \cite{Yuneec}, and Autel Evo II series \cite{Autel}. Previous studies \cite{son2015rocking, davidson2016controlling, nassi2019drones} investigated the security of drones. They either launched DoS attacks on the drone by injecting ultrasound into the IMU \cite{son2015rocking, wang2017sonic, yan2020surfingattack} in a close attack range (around $10 cm$), or aimed at controlling the drone in an indoor environment with a limited attack range ($\leq 3 m$) \cite{davidson2016controlling}. \emph{“How to control the drone over a long range”} is still an open problem. 
In this work, we demonstrate the capability of \attack in controlling 
drones continuously over a long range. As shown in Fig. \ref{pic1}, an attacker uses two projectors to launch \attack at 7$m$ away on a flying DJI drone. In doing so, we expose the new threats against the stereo cameras in OA systems.

\begin{figure*}[t!]
	\centering
	\subfloat[Human binocular vision]{\includegraphics[trim={2mm 20mm 8mm 2mm},clip,width=0.3\textwidth]{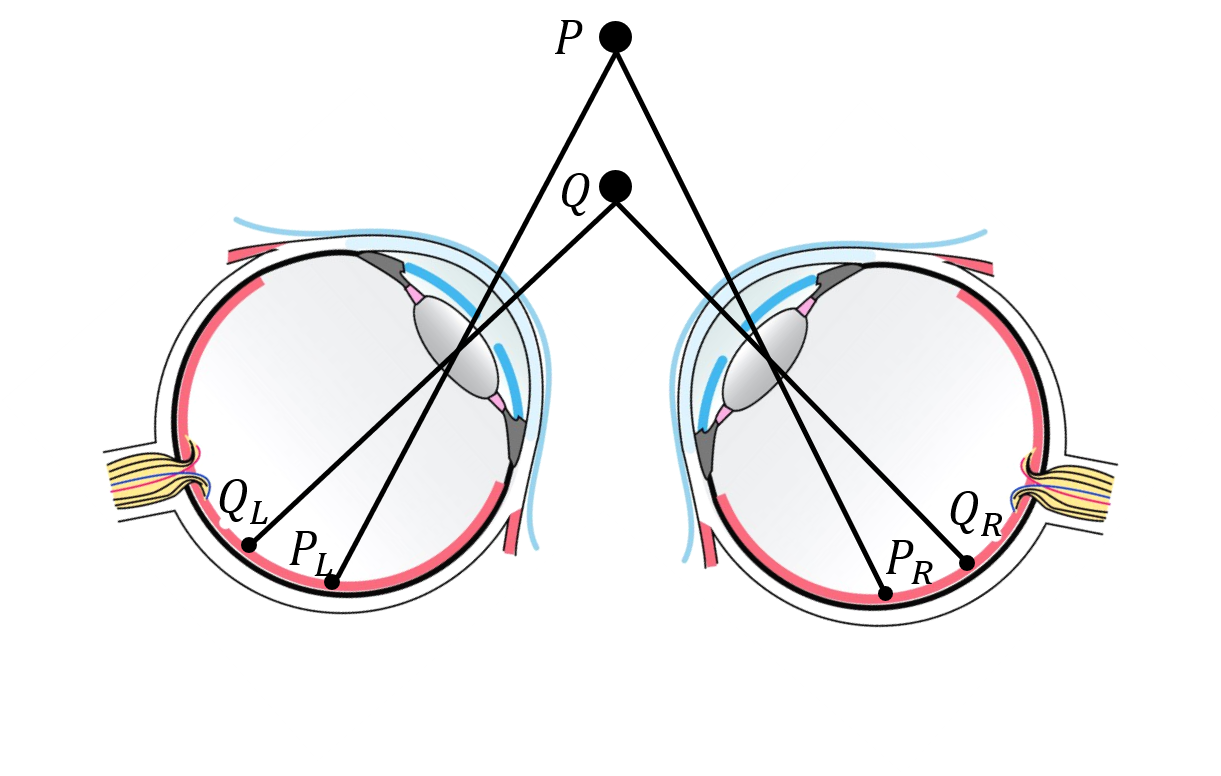}
		\label{pic2a}}
	\subfloat[Camera stereo vision]{\includegraphics[width=0.3\textwidth]{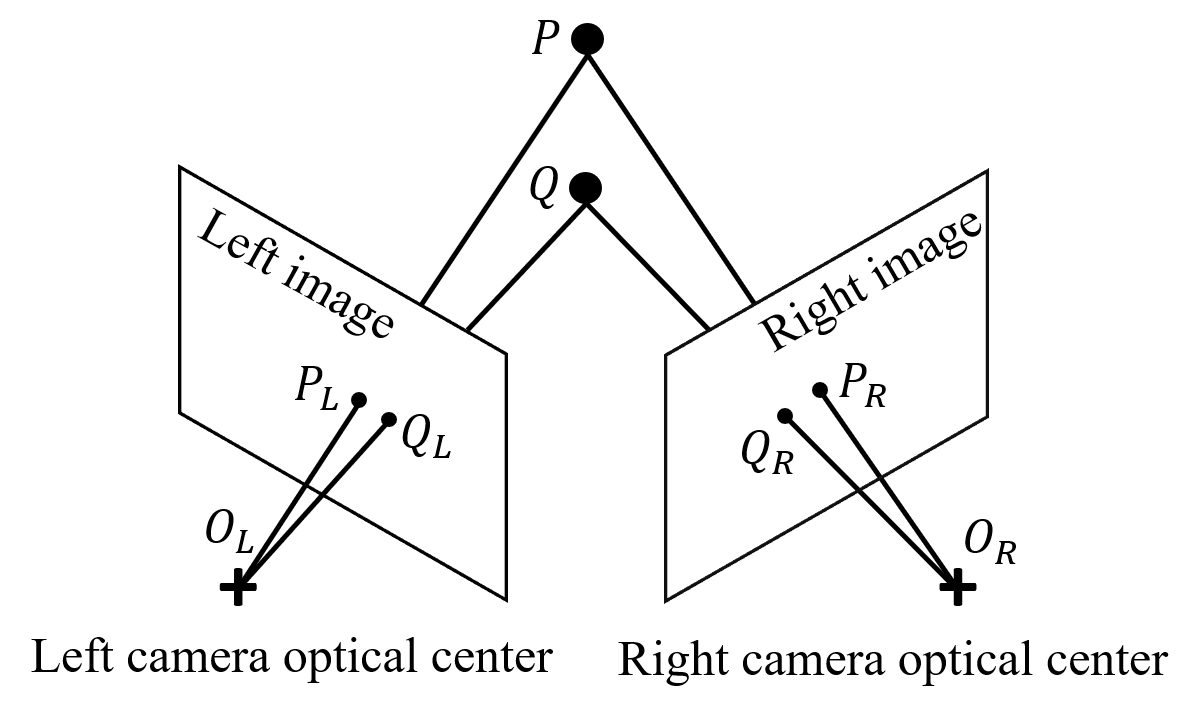}
		\label{pic2b}}
	\subfloat[Triangulation in depth estimation]{\includegraphics[width=0.3\textwidth]{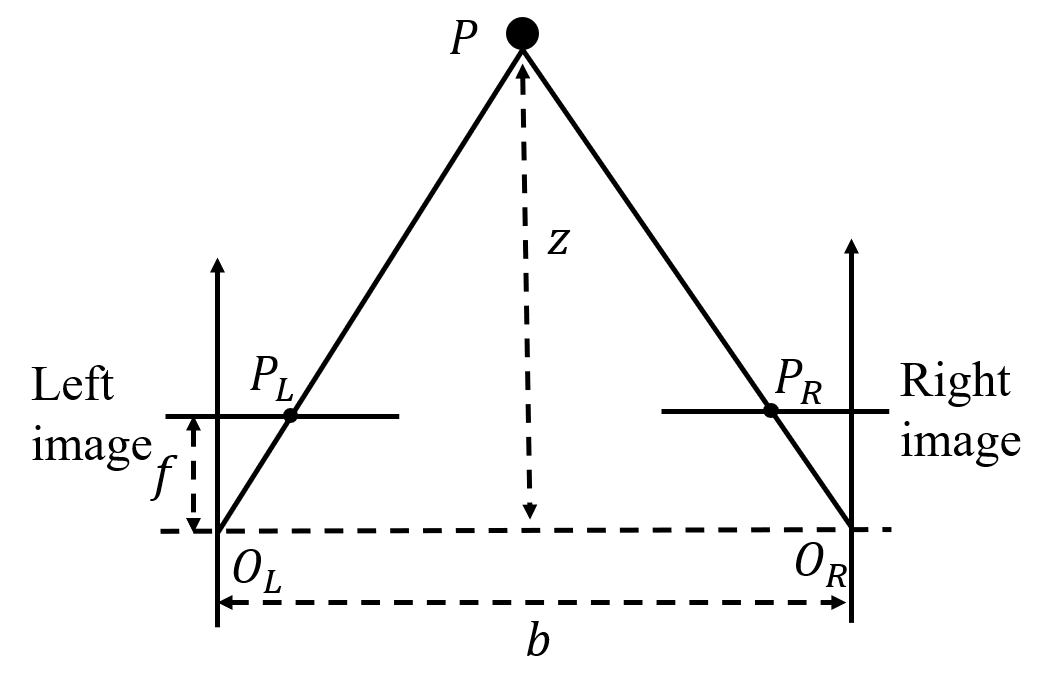}
		\label{pic2c}}
	\caption{Depth estimation in stereo vision.}
\end{figure*}

\attack consists of two attack formats: the \emph{beams attack} and \emph{orbs attack}. They mainly exploit the false stereo matching in depth estimation algorithms and the lens flare effects in optical imaging. 
\emph{Beams attack} exploits the stereo matching process, in which an attacker injects two different light sources into each camera. The injected light source will become brighter and more prominent in one camera than the other. Since stereo matching tries to find the pixels in images corresponding to the same point, it recognizes the injected strong light sources in left and right images as the same object, resulting in a faked object depth. 
\emph{Orbs attack} leverages lens flare effect, a phenomenon that strong light beams are refracted and reflected multiple times, which creates green-color polygon-shape artifacts in camera images~\cite{Prevent_flare, hullin2011physically}. 
When two light sources are injected into each camera, a green orb will be created for each image. The depth estimation algorithms falsely match two orbs in two images as the same object, resulting in a fake obstacle depth.  

There are two major challenges in realizing \attack: (1) How to design the attacks that can induce
variable fake obstacle depths? (2) How to estimate the artificial obstacle’s position without accessing the stereo camera? To address these challenges, we design 3 different attack patterns for both the orbs and beams attacks, totaling 6 attack patterns. We launch different types of attacks in tandem to complement each other in extending the attack range while maintaining high attack success rate.

We verify the efficacy of \attack both in simulation and in real-world experiment. We run the simulation on one of the most popular unmanned vehicle projects, Ardupilot \cite{Ardupilot}, to demonstrate the potential attacks on drones and vehicles. For the proof-of-concept experiments, we conduct our attacks on two commercial stereo cameras designed for autonomous systems (e.g., robotic vehicles, drones, robots), ZED \cite{ZED} and Intel RealSense D415 \cite{Intel}, and one popular drone, DJI Phantom 4 Pro V2 \cite{DJI_Phantom_4}. Evaluation results show that our attacks can achieve up to 15$m$ distance at night and up to 8$m$ distance during daytime with fake obstacle distance ranging from 0.5$m$ to 16$m$, which covers the whole range of the obstacle depth in the OA system on the DJI drone. Both the simulation and physical world experiments demonstrate that the  devices under attack run out of control as soon as our attacks are turned on. We set up a website\footnote{\url{https://fakedepth.github.io/.}} to show the simulation results and demo videos. 

In summary, this paper makes the following contributions:
\begin{itemize}
   \item We propose \attack, the first attack against stereo vision-based depth estimation in OA systems on robotic vehicles and drones. 
   \item We are the first to launch a \emph{long-range and continuous} attack towards autonomous systems. Through simulation, we show that \attack achieves a fine-grained trajectory manipulation of drones. 
    \item We successfully launch \attack on two commercial stereo cameras designed for robotic vehicles and drones (ZED and Intel RealSense D415) and one of the most advanced drones (DJI Phantom 4 Pro V2) in different ambient light conditions.
\end{itemize}

\section{Background}\label{background}

\renewenvironment{comment}{}{
\begin{figure*}[t]
\centering
\subfloat[X-shape beams attack from 9$m$ away ]{\includegraphics[trim={3.5mm 0mm 0mm 0mm},clip,width=\textwidth]{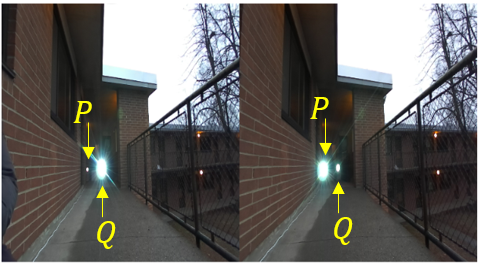}
\label{pic4a}}

\subfloat[Trapezoid-shape orbs attack from 2$m$ away ]{\includegraphics[width=\textwidth]{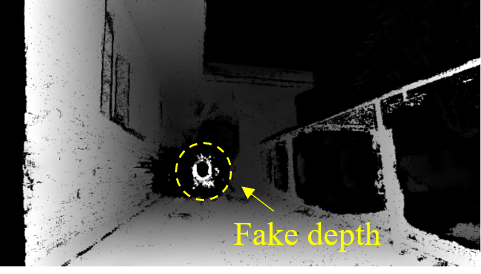}
\label{pic4b}}
\caption{Beams attack and orbs attack on ZED stereo camera during the daytime. From left to right in (a) and (b), they are left and right stereo camera images, depth map, and 3D point cloud. In the depth map, 100\% white means the closest distance and black means the furthest. 3D point cloud is the reconstruction of the 3D scene based on the depth map. Both (a) and (b) show the visualization of fake depths in the depth map and their corresponding fake obstacle in the 3D point cloud.}
\label{pic4}
\end{figure*}
}

In this section, we briefly introduce the preliminary background knowledge of \attack, including the foundation of depth estimation from stereo vision and the lens flare effect in optical imaging. 

\subsection{Depth Estimation from Stereo Vision}

Stereo vision-based depth estimation transforms 2D images into the perception of the 3D depth map, which contains information related to the distance of the scene objects from the stereo camera \cite{Stereoscopy}. Depth estimation algorithms are mostly used in autonomous systems to detect and avoid obstacles. The main idea of depth estimation using a stereo camera involves the concept of stereo matching and triangulation.

In essence, stereo camera resembles human binocular vision shown in Fig. \ref{pic2a}, which is composed of two lenses with a separate image sensor for each lens, \textit{i.e.,} left image and right image in Fig. \ref{pic2b} \cite{Stereo_camera}. Binocular disparity in Fig. \ref{pic2a} refers to the difference in the image location of an object seen by the left and right eyes, resulting from the eyes' horizontal separation. Binocular disparity is the basis for extracting depth information from 2D retinal images in stereopsis \cite{Binocular_disparity}.
For example, in Fig. \ref{pic2a}, $P_L$ and $P_R$ represent the corresponding images of $P$ shown on the left and right retinas, while $Q_L$ and $Q_R$ are the corresponding images of $Q$. The binocular disparities introduce a difference in the sensed depth of $P$ and $Q$. Likewise, the disparity in the stereo camera refers to the difference in the coordinates of similar features in two images, e.g., $P_L$ and $P_R$ in Fig. \ref{pic2b}.

Triangulation method is used by stereo vision to calculate depth \cite{George_Bebis}.
For example, in Fig. \ref{pic2c}, $O_L$ and $O_R$ represent the left and right optical centers in the stereo camera.
The intersection points $P_L$ and $P_R$ are on the left and right images, respectively.
The depth of point $P$ is calculated using similar triangles, $\triangle{P_{L}PP_{R}}$ and $\triangle{O_{L}PO_{R}}$, donated as $\triangle{P_{L}PP_{R}} \sim \triangle{O_{L}PO_{R}}$.
Suppose the horizontal axis values of $P_{L}$ and $P_{R}$ are $p_{l}$ and $p_{r}$, respectively. Since the ratio of corresponding sides is equal in similar triangles, we have: 
\begin{equation}\label{eq1}
    \frac{b}{z}=\frac{b+p_{r}-p_{l}}{z-f},
\end{equation}
where $z$ is the depth of point $P$, $f$ is the focal length of the camera, $b$ is the baseline (distance between the two camera optical centers), $p_{l}-p_{r}$ is the disparity between $P_{L}$ and $P_{R}$. Therefore, we can derive the depth of point $P$ as: 
\begin{equation}\label{eq2}
    z = \frac{f\cdot b}{p_{l}-p_{r}}. 
\end{equation}
Triangulation relies on the stereo matching, which establishes the pixel correspondence between primitive factors in images, e.g., $P_{L}$ in the left image corresponds to $P_{R}$ in the right image in Fig.~\ref{pic2b}. Once the stereo correspondence is established, we can compute each pixel's disparity in the images to form a disparity map, which can be converted into a 3D depth map using Eq.~(\ref{eq2}). However, certain interference factors, such as illumination, noise, and surface physical characteristics, could induce ambiguous correspondence between points in the two images, e.g., the depth estimation algorithm falsely takes $P_{L}$ in the left image as the correspondence of $Q_{R}$ in the right image in Fig.~\ref{pic2b}. Such ambiguous correspondence caused by stereo mismatching may lead to inconsistent interpretations of the same scene, aggravating the false depth calculation.  

\begin{figure}[b]
\centering
\begin{minipage}{0.5\textwidth}
  \centering
\subfloat[]{\includegraphics[trim={5mm 0mm 0mm 5mm},clip,width=0.42\textwidth]{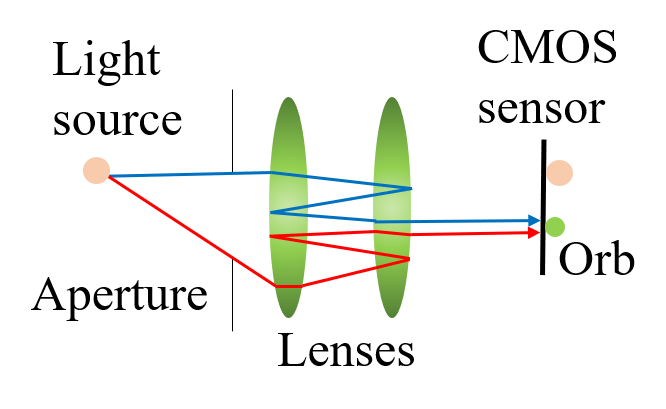}
\hspace{1mm}
\label{pic3a}}
\subfloat[]{\includegraphics[width=0.42\textwidth]{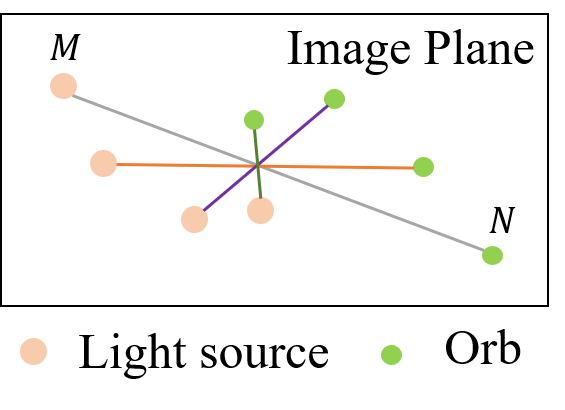}
\label{pic3b}}
\end{minipage}
\caption{(a) Lens flare effect; (b) the relationship between the light source and the orb's position.
} 
\label{pic3}
\end{figure}

\begin{figure*}[!t]
    \centering
    \subfloat[Left and right images]{\includegraphics[trim={2mm 0mm 2mm 0mm},clip,width=0.27\textwidth]{figures/Picture4a.png}\label{pic4a}}
    \hspace{2mm}
    \subfloat[Depth map]{\includegraphics[trim={0mm 0mm 1mm 0mm},clip,width=0.27\textwidth]{figures/Picture4b.png}\label{pic4b}}
    \hspace{2mm}
    \subfloat[3D point cloud]{\includegraphics[trim={1mm -1mm 2mm 0.8mm},clip,width=0.27\textwidth]{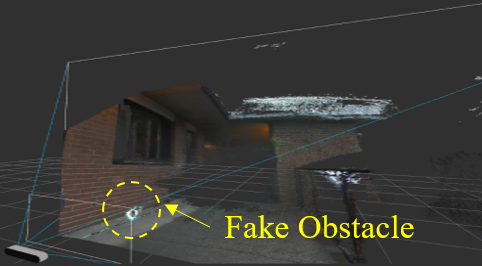}\label{pic4c}}\\
    \subfloat[Left and right images]{\includegraphics[width=0.27\textwidth]{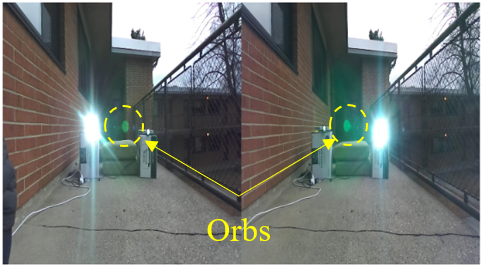}\label{pic4d}}
    \hspace{2mm}
    \subfloat[Depth map]{\includegraphics[trim={1mm 0mm 1mm 0mm},clip,width=0.27\textwidth]{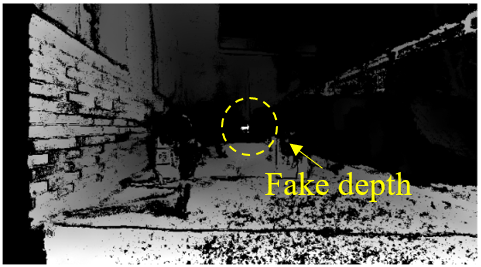}\label{pic4e}}
    \hspace{2mm}
    \subfloat[3D point cloud]{\includegraphics[width=0.27\textwidth]{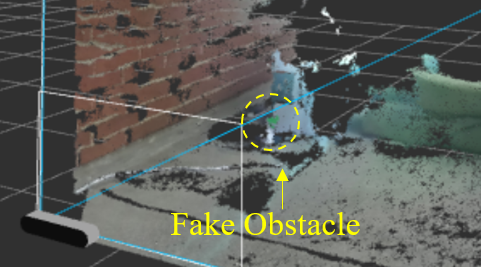}\label{pic4f}}
    \caption{\attack on ZED stereo camera during the daytime. (a) to (c) shows the example of X-shape beams attack from 9$m$ away, and (d) to (f) showcase the trapezoid-shape orbs attack from 2$m$ away. In the depth map, 100\% white means the closest distance and black means the furthest. The 3D point cloud is the reconstruction of the 3D scene based on the depth map.}
    \label{pic4}
\end{figure*}

\renewenvironment{comment}{}{
\begin{figure}[b!]
\centering
\subfigure[]{\includegraphics[trim={5mm 0mm 0mm 5mm},clip,width=0.2\textwidth]{figures/Picture3a.png}
\hspace{1mm}
\label{}}
\subfigure[]{\includegraphics[width=0.2\textwidth]{figures/Picture3b.png}
\label{}}
\caption{(a) Lens flare effects (b) The relationship between the light source and the orb's position} 
\label{pic3}
\end{figure}
}

\subsection{Lens Flare Effect}


Lens flare effect \cite{hullin2011physically, man2020ghostimage} 
is a phenomenon caused by the scattering and reflection of a bright light through a non-ideal lens system, where several undesirable artifacts appear on an image (Fig. \ref{pic3a}). 
Ideally, all light beams should pass directly through the lens to reach the image sensor, and they will form a bright glare in the image. 
However, due to lens imperfections, a small portion of the light will be reflected several times within the lens system before reaching the image sensor. Such reflections will result in multiple artifacts on the image. Under normal light conditions, these artifacts are usually invisible. However, when a strong light source (e.g.,  sun, light bulb, projector) is present, the lens flare becomes more visible and washes out the entire image.

Anti-reflection coatings on the lenses of the commercial cameras are often used to mitigate the lens flare effect by filtering out most reflections \cite{Canon_Lab}, but these lenses still suffer from the green-color flare orb if a strong white-light source is present. Note that, according to our experiments, the relationship between the green-color orb's position and the white-light source is centrosymmetric as illustrated in Fig. \ref{pic3b}\footnote{\rev{In crystallography, a centrosymmetric point group contains an inversion center as one of its symmetry elements. 
For example, points $M$ and $N$ in Fig. \ref{pic3b} are centrosymmetric to each other, i.e., if the coordinate of $M$ in the 2D image is $(a,b)$, the coordinate of $N$ will be $(-a,-b)$.}}. 
In fact, most of commercial cameras nowadays have applied the anti-reflection coatings \cite{FindLight}. 

\section{Vulnerability in Depth Perception}\label{vulnerability}

\begin{figure*}[t]
	\centering
	\subfloat[X-shape attack]{\includegraphics[width=0.3\textwidth]{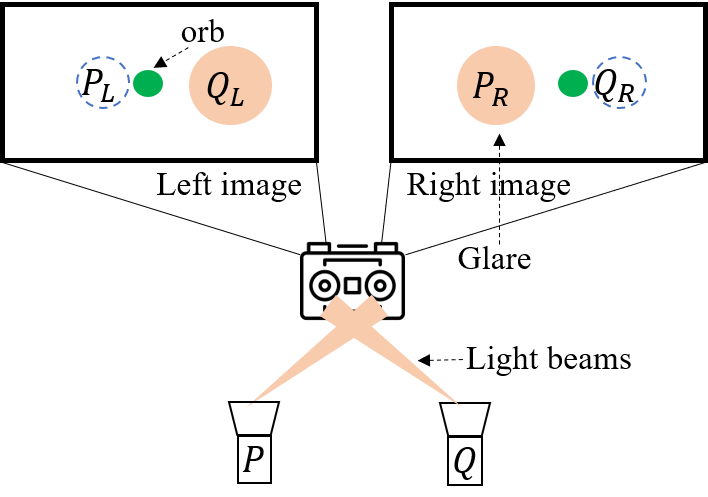}
	\hspace{2mm}
		\label{pic5a}}
	\subfloat[Trapezoid-shape attack]{\includegraphics[width=0.3\textwidth]{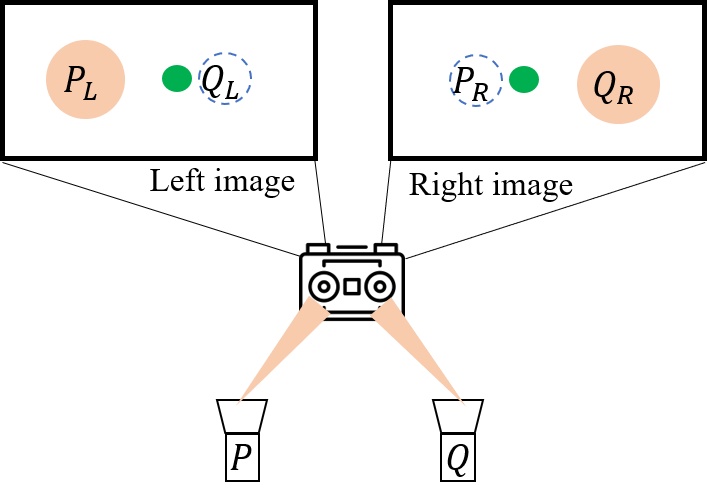}
	\hspace{2mm}
		\label{pic5b}}
	\subfloat[Triangle-shape attack]{\includegraphics[width=0.3\textwidth]{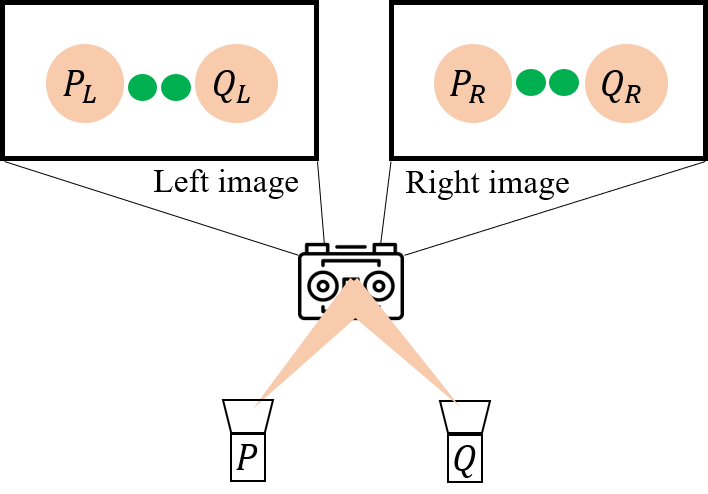}
	\hspace{2mm}
		\label{pic5c}}
	\caption{Three different attack patterns are used in beams and orbs attacks. The orange circle represents the appearance of the injected light (glare) in the image, and the dotted white circle stands for the position of the light source which is not aiming at the camera. Green small circles stand for the orbs.}
	\label{pic5}
\end{figure*}

In this section, we first present the vulnerabilities of depth perception in computer vision exploited by \attack, and then we illustrate the main experimental observations that lead to the design of \attack.

The main root cause of the vulnerabilities in depth estimation algorithms is their lack of higher-order perceptions. Unlike human perception, such algorithms do not base their decisions on personal experience. 
Instead, they aim to match similar features (e.g., shapes, colors) with high confidence as stereo correspondence, as long as they are in relatively reasonable positions in the left and right images.
Therefore, most depth estimation algorithms can be affected by physical perturbations.
Our experiments demonstrate that strong light beams and lens flare orbs can induce wrong depth perception due to the \emph{light beams mismatching} and \emph{false orbs matching}.

\textbf{Light Beams Mismatching.} As shown in Fig. \ref{pic4a}, an attacker injects the light beams with the same projection intensity using two projectors (\emph{i.e.,} $P$ and $Q$) into the right and left cameras, respectively. It can be observed that one of the injected light sources is brighter than the other one when received by the camera, which becomes a more prominent feature in the image. For example, $Q$ is targeting the left camera, so the light source shown on the left image from $Q$ is brighter than that from $P$. Such phenomenon causes the depth estimation algorithms to mismatch the two highlighted light sources in the images as the same object due to the aforementioned weakness in depth perception. 
As a result, a fake near-distance depth is created in the depth map (Fig. \ref{pic4b}) and visualized in the 3D point cloud (Fig. \ref{pic4c}). By adjusting the distance between projectors and the stereo camera, as well as the distance between two projectors, different fake obstacle depths can be created.

\textbf{False Orbs Matching.} Fig. \ref{pic4d} shows an example that the two orbs generated by the strong light sources in the stereo images can mislead the depth perception to falsely identify them as a 3D obstacle. By matching them as the stereo correspondence of one another, the targeted depth estimation algorithm outputs a fake depth in depth map (Fig. \ref{pic4e}) and a corresponding 3D fake obstacle in 3D point clouds (Fig. \ref{pic4f}). Due to the centrosymmetry of the light source and the orb, the attacker is able to adjust the angle of the injected light to control the orb's position. Exploiting such a phenomenon, theoretically, an attacker could adjust the positions of the two orbs in the stereo camera to manipulate the fake depth values.

\section{Threat Model}\label{threat_model}



In this section, we present the threat model of this work, including the attack goal, the attacker's capability, and the attack scenarios.
The attacker's goal is to disrupt regular operations of autonomous systems by injecting fake depth, 
and further lead to unintended system behaviors.
For example, 
an attacker can force a drone into a severe crash, e.g., hitting a tree, by changing the depth of a real obstacle. 


Our attack is a \emph{fully black-box} attack against general stereo vision based depth estimation algorithms used in OA systems. An attacker has no physical access to the hardware or firmware of the attack target, nor does he/she have access to the camera images. An attacker also has no prior knowledge about the depth estimation algorithms used in the OA system. 

We consider an attack target is equipped with a stereo camera for OA. 
For drone attacks, we further assume the drones operate in a flying mode, such as Positioning (P) Mode in DJI drones \cite{DJIUserManual} or Loiter Mode in Ardupilot \cite{Ardupilot}, where a human operator controls the drone with the assistance of the OA. However, once an obstacle is detected within its OA range, the OA in the autonomous system makes decisions preceding any human input, \emph{i.e.}, it takes precedence over the pilot/driver. As an example, when an obstacle is detected in front of the drone within the safety margin of OA, the drone will stop moving forward even if the pilot pushes forward the throttle.

\begin{figure*}[t!]
\centering
\subfloat[Beams attack: $b > d$]{\includegraphics[width=0.23\textwidth]{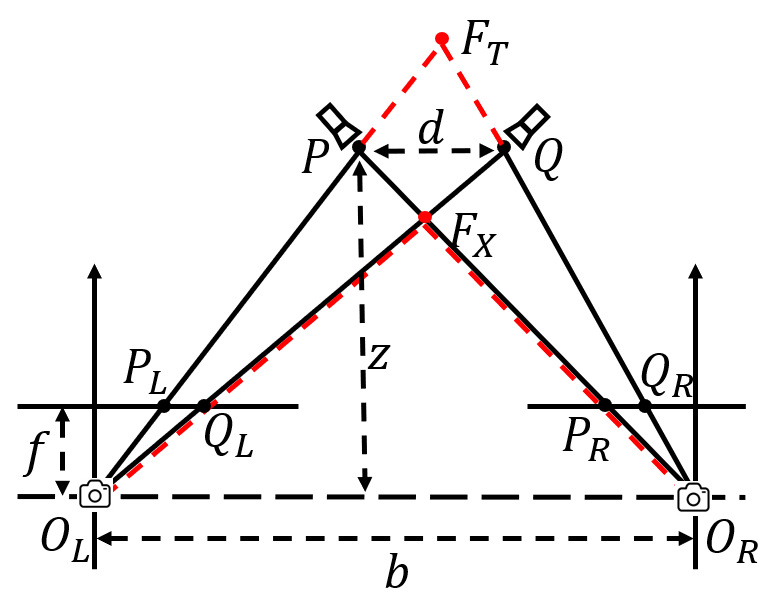}
\label{pic6a}}
\subfloat[Beams attack: $b < d$]{\includegraphics[width=0.22\textwidth]{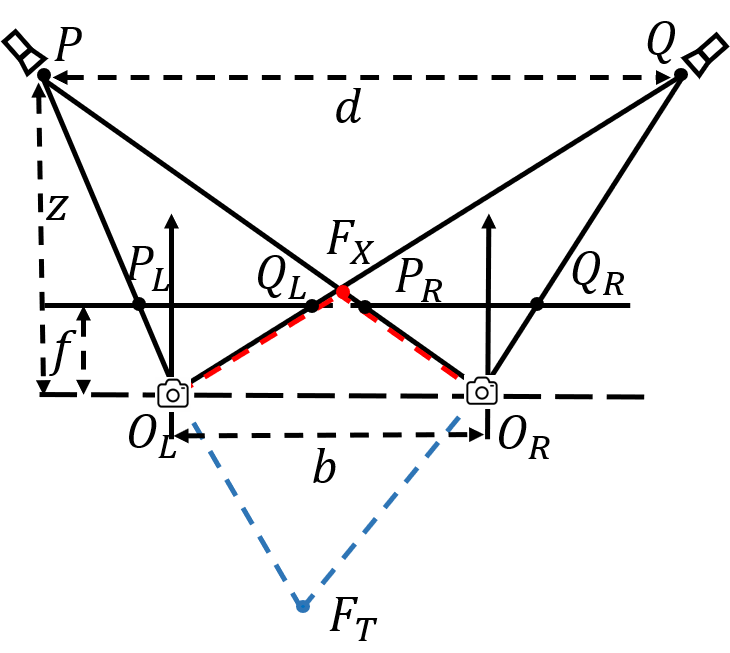}
\label{pic6b}}
\subfloat[Orbs attack: $b > d$]{\includegraphics[width=0.3\textwidth]{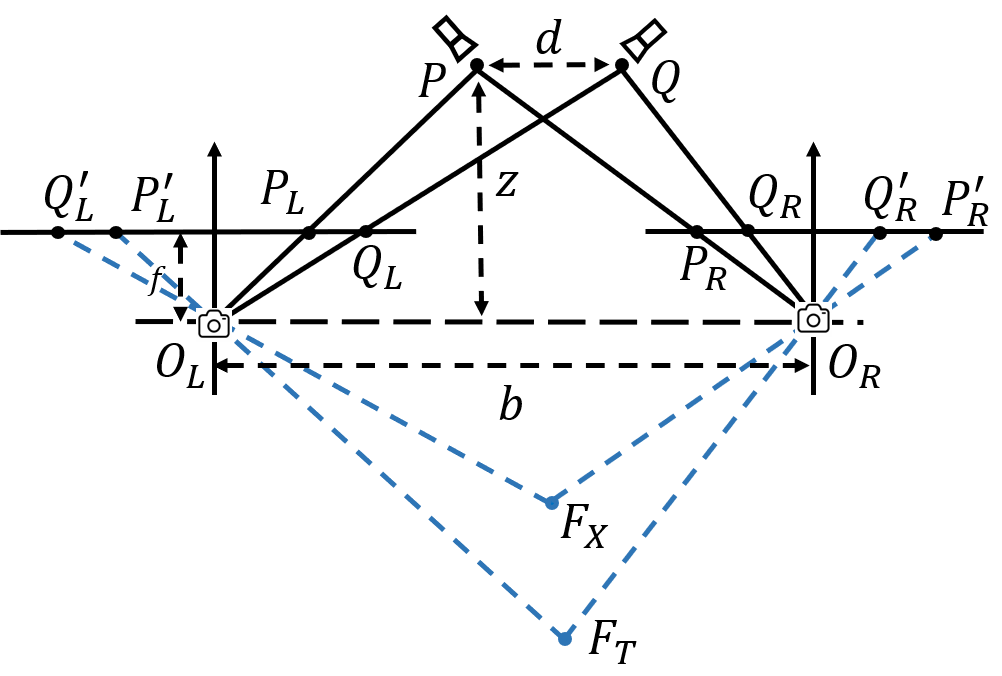}
\label{pic6c}}
\subfloat[Orbs attack: $b < d$]{\includegraphics[width=0.23\textwidth]{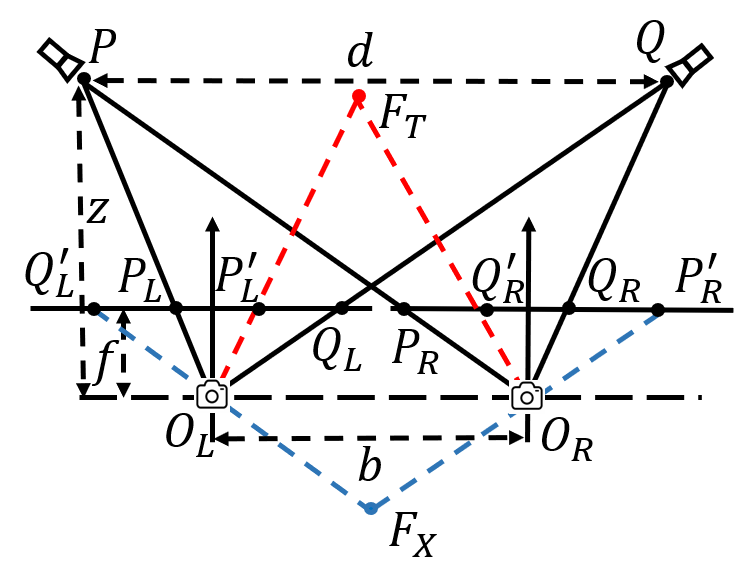}
\label{pic6d}}
\caption{Obstacle positions in beams attack (a-b) and orbs attack (c-d) in two attack scenarios. $b$ is the baseline of the stereo camera, and $d$ is the distance between the two light sources. $P$ and $Q$ represent the positions of the two light sources, and $O_L$ and $O_R$ represent the left and right optical centers of the stereo camera. $z$ is the true depth of the light sources, corresponding to the perpendicular distance between the projectors and the stereo camera. $P_L$ and $Q_L$ are the intersections of $P$ and $Q$ on the left image, while $P_R$ and $Q_R$ are the intersections on the right image. $Q_{L}'$, $Q_{R}'$, $P_{L}'$ and $P_{R}'$ are the location of the orbs. $F_X$ and $F_T$ represent the fake obstacle positions corresponding to the X-shape and trapezoid-shape attack, where the red color implies the existence of fake obstacles while blue implies their non-existence in theory.
} \label{pic6}
\end{figure*}
\section{DoubleStar} \label{attack_design}

This section introduces the design of \attack, including the beams attack and orbs attack, and analyzes these attacks via 
mathematical models.  

\subsection{Attack Overview}
\attack exploits the vulnerabilities in the depth perception. Certain interference factors can cause ambiguous correspondences between points in the two images, which can lead to alternative interpretations of the scene. We design the beams attack using two light sources to form glares on the images which can cause false stereo correspondence in depth estimation algorithms and further lead to fake depth creation. Besides, when a light source is targeting the camera, the lens flare effect will bring in an orb in the image which is centrosymmetric to the injected light source. The orbs attack mainly utilizes this phenomenon to fool the depth estimation algorithms by matching the two generated orbs in two images as the stereo correspondence. 
Fig. \ref{pic5} presents the design of the beams attack and orbs attack with three attack patterns. $P$ and $Q$ are the light sources, e.g., projectors. Their corresponding position in the left and right image in the stereo cameras are $P_L$, $Q_L$, $P_R$, $Q_R$, respectively. A green orb is centrosymmetric to the injected light source in each image.

We design three different patterns for both beams and orbs attacks, 
as shown in Fig. \ref{pic5}: 1) \emph{\textbf{X-shape attack}} (Fig. \ref{pic5a}): $P$ is pointing to the right camera while $Q$ is aiming at the left camera. \emph{\textbf{Trapezoid-shape attack}} (Fig. \ref{pic5b}): $P$ is targeting the left camera whereas $Q$ is pointing to the right camera.
\emph{\textbf{Triangle-shape attack}} (Fig. \ref{pic5c}): $P$ and $Q$ are covering the left and right cameras at the same time. 
Here, we define the \emph{depth of near-distance fake obstacles} as the one smaller than the true depth of the light source, and define the \emph{depth of far-distance fake obstacles} as the one greater than the true depth. 

\subsection{Beams Attack}
\subsubsection{Attack Design}

\textbf{X-shape Attack.} Using X-shape attack pattern (Fig. \ref{pic5a}), $P$ and $Q$ form the corresponding glares $P_R$ and $Q_L$, respectively. With certain constraints on the perpendicular distance and included angle between the light sources and the stereo camera, the targeted depth estimation algorithm falsely takes $P_R$ as the stereo correspondence of $Q_L$. As a result, it outputs a near-distance fake depth since the disparity becomes larger than the real one. 
Fig. \ref{pic4a} illustrates a real-world scenario of the X-shape beams attack which produces a near-distance fake depth (Fig. \ref{pic4b}), when the real depth of the projectors are 9$m$ away from the stereo camera. 

\textbf{Trapezoid-shape Attack.} As shown in Fig. \ref{pic5b}, by using the trapezoid-shape attack pattern, the glares $P_L$ and $Q_R$ are generated by $P$ and $Q$. Similarly, with certain constraints on the distance and angle, the depth estimation algorithm matches $P_L$ and $Q_R$ as the same object. However, it only works when the glares approach the center of the image, otherwise, no fake depth can be generated, the reason of which is explained in Section \ref{math_beams}. Since the glares cover multiple pixels, the algorithm will output a far-distance fake depth, overshadowing the real obstacle depth. The generated fake depth merges into the background depth, which cannot be identified by human eyes. However, we confirm that the fake depth can be perceived by the stereo cameras. 

\textbf{Triangle-shape Attack.} Fig. \ref{pic5c} shows the triangle-shape beams attack where the two light beams cover both the left and right cameras. Hypothetically, when the injected light intensities from $P$ and $Q$ reflected on both left and right images are equal to each other, the depth estimation algorithm will match $P_L$ and $P_R$ as the same object. Similarly, $Q_L$ and $Q_R$ will also be matched as the stereo correspondence. Thus, the algorithm outputs the true depths of $P$ and $Q$. However, in a real-world attack,
when the injected light intensities on the left and right cameras slightly differ, the triangle-shape attack will be transformed into the X-shape or trapezoid-shape attack. 

\subsubsection{Mathematical Modeling and Analysis}\label{math_beams}

Since the triangle-shape attack is essentially the X-shape or trapezoid-shape attack, we conduct a mathematical analysis of these two most basic attack patterns. 
Figs.~\ref{pic6a} and \ref{pic6b} present the mathematical model for beams attacks. 

Suppose the depth of $F_X$ and $F_T$ are $z_x$ and $z_t$, respectively. The corresponding coordinates of $P_L$, $P_R$, $Q_L$ and $Q_R$ are $p_l$, $p_r$, $x_l$ and $x_r$. Based on Eq. (\ref{eq2}), the disparity of point $P$ is: 
    $p_{l}-p_{r} = \frac{f\cdot b}{z}$.
Moreover, since $\triangle{O_{L}QP} \sim \triangle{O_{L}Q_{L}P_{L}}$, we have:
    $\frac{z}{f}=\frac{d}{q_{l}-p_{l}}$,
    $q_{l}=\frac{f\cdot d}{z}+p_{l}$.
The disparity of $F_X$ is: 
\begin{equation} \label{eq3}
    \begin{split}
        q_{l}-p_{r}=\frac{f\cdot d}{z}+p_{l}-p_{r}=\frac{f\cdot (d+b)}{z}.
    \end{split}
\end{equation}
Thus, from the Eqs.~(\ref{eq2}) and (\ref{eq3}), we obtain the fake depth $z_x$:
\begin{equation}\label{beam_x}
    z_{x}=\frac{b}{d+b}\cdot z.
\end{equation}
Obviously, 
$z_x < z$, which indicates that the fake obstacle $F_X$ created by the X-shape attack is nearer to the stereo camera than the light sources as shown in Figs. \ref{pic6a} and \ref{pic6b}. Note that when $0 < z_x < f$, $F_X$ is non-existent. This follows the optical imaging principle that the fake depth cannot be shorter than the focal length $f$. Such scenario appears either when the two projectors are too far away from each other or the perpendicular distance between the projectors and the stereo camera is too small. In other words, if $d$ is too large or $z$ becomes too small, the fake obstacles may not be created. The analysis of a failed attack scenario can be found in the Appendix \ref{unsuccess_case}.

Similarly, $z_t$ can be expressed as:
\begin{equation}\label{beam_t}
    z_{t}=\frac{b}{b-d}\cdot z,
\end{equation}
where $z_t > z$ if $b > d$, and $z_t < 0$ if $b < d$. Correspondingly, as shown in Fig. \ref{pic6a}, when $b > d$, $F_T$ has a larger depth. Conversely, when $b < d$, $F_T$ appears on the opposite side of the stereo camera, which is non-existent. However, the injected light is not a single pixel, instead, it contains several blocks with multiple pixels in the image. The depth estimation algorithm will try to match these blocks in the left and right images,
which could result in a far-distance fake depth. This special case is marked in blue in Fig. \ref{pic6b}. 

\subsection{Orbs Attack}
\subsubsection{Attack Design}

\textbf{X-shape Attack.} Fig. \ref{pic5a} shows that the generated green orbs are centrosymmetric to the glares $Q_L$ and $P_R$ in the left and right images (see Appendix~\ref{green_orbs}). Given proper attack distance/angle, the depth estimation will falsely match the two orbs as the same object. However, the attack works only when the orbs approach the image center, otherwise, no fake depth is created (see Section \ref{math_orbs} for the reason). Since orbs consist of multiple pixels, the algorithm can output a far-distance fake depth, whose exact value depends on the orbs' positions.

\textbf{Trapezoid-shape Attack.} Since the orb and the glare are centrosymmetric, the orb appears at the right of the $P_L$ in the left image, and at the left of the $Q_R$ in the right image (Fig. \ref{pic5b}). As mentioned before, due to the weakness of the depth perception, the depth estimation algorithm matches the two orbs as the same object and outputs a near-distance fake depth since the disparity is larger than the real one. Fig. \ref{pic4d} shows that a real-world trapezoid-shape orbs attack is able to create a near-distance fake depth (Fig. \ref{pic4e}) by matching the two orbs in the left and right images as the stereo correspondence. 

\textbf{Triangle-shape Attack.} Since the projection from $P$ and $Q$ covers both cameras, four glares $P_L$, $Q_L$, $P_R$, and $Q_R$ appear in the left and right images. Based on the centrosymmetric relationship between the glare and orb, two orbs appear in each image. The depth estimation matches the orbs in the left and right image correspondingly. Due to the centrosymmetry, the depth estimation outputs two fake obstacle depths that are the same as the light source's real depth. However, in practice, since the attacker cannot precisely control every single pixel in the image, 
the stereo correspondence will occur either like Fig. \ref{pic5a} or Fig. \ref{pic5b}. As a result, the fake depth is still created. 

\renewenvironment{comment}{}{
\begin{figure}[t!]
\centering
\begin{minipage}{0.5\textwidth}
  \centering
\subfloat[$b > d$]{\includegraphics[width=0.7\textwidth]{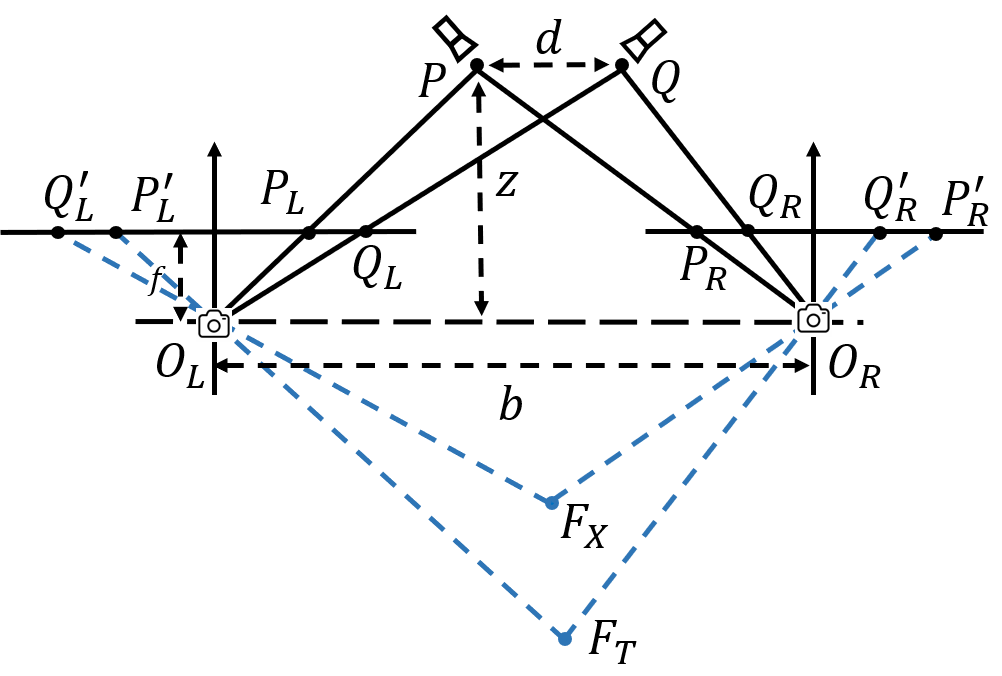}
\label{pic7a}}

\subfloat[$b < d$]{\includegraphics[width=0.55\textwidth]{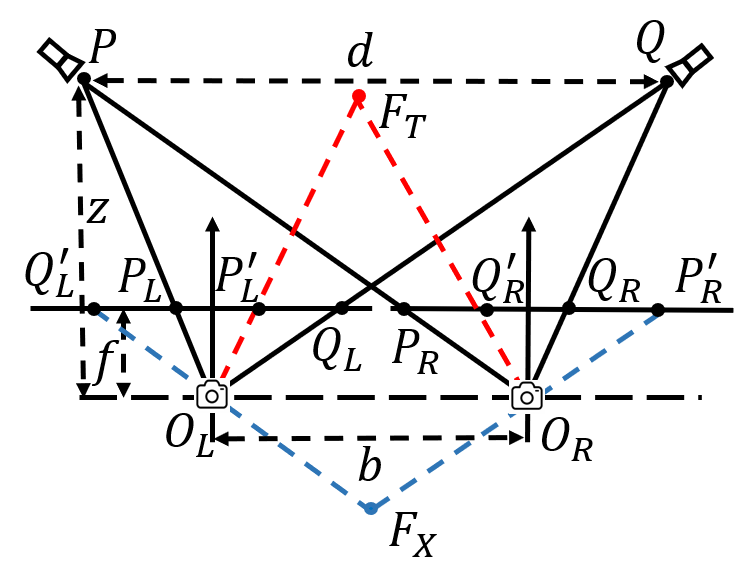}
\label{pic7b}}
\end{minipage}
\caption{Obstacle positions in orbs attack under different relationship between $d$ and $b$. Red and blue stands for existent and non-existent fake obstacle, respectively.} \label{pic7}
\end{figure}
}

\subsubsection{Mathematical Modeling and Analysis}\label{math_orbs}

Figs. \ref{pic6c} and \ref{pic6d} show the mathematical modeling of orbs attack. The corresponding horizontal coordinates of the orbs $Q_{L}'$, $Q_{R}'$, $P_{L}'$ and $P_{R}'$ are $q_{l}'$, $q_{r}'$, $p_{l}'$ and $p_{r}'$, respectively. In the orbs attack, we also consider two scenarios based on the relationship between $b$ and $d$.

Since the orbs and their corresponding injected lights are centrosymmetric, $p_{l}'$, $p_{r}'$, $x_{l}'$ and $x_{r}'$ have the exact opposite value as $p_{l}$, $p_{r}$, $x_{l}$ and $x_{r}$, respectively. Following the  derivation in Section \ref{math_beams}, $z_x$ and $z_t$ for orbs attack are: 
\begin{equation}\label{orb_x}
    z_{x}=-\frac{b}{d+b}\cdot z,
\end{equation}
\begin{equation}\label{orb_t}
    z_{t}=\frac{b}{d-b}\cdot z.
\end{equation}
Since 
$z_x < 0$, it means that $F_X$ is non-existent regardless of the value of $d$. For the depth of $F_T$, when $b < d$, 
$z_t < z$, which indicates that a near-distance depth can be created by trapezoid-shape attack in Fig. \ref{pic6d}. When $b > d$, $z_t < 0$, $F_T$ appears on the other side of the stereo camera indicating the non-existence of the $F_T$.
 However, since the orbs consist of multiple pixels, it is still possible for a far-distance fake depth to be formed. The three special cases are marked in blue. 

It is worth noting that when $d = b$, no fake depth can be generated in both attacks. In summary, (1) the beams attack works with all three attack patterns when $b > d$, and with X-shape and triangle-shape attack pattern when $b < d$; (2) the orbs attack works with trapezoid-shape and triangle-shape attack when $b < d$. Comparing the blue and red fake obstacle points in beams and orbs attack (Fig. \ref{pic6}), we can see that the beams and orbs attacks complement each other's performance. As a result, in a real-world attack, fake obstacles generated by beams attack and orbs attack can co-exist, and these two attacks can operate in concert to enhance the attack capability. 


\section{Simulation} \label{simulation}

\minor{In this section, we first evaluate \attack against drone in a simulation environment. Then, we simulate the attacks towards various stereo depth estimation algorithms to verify the attack impact.}

\subsection{Drone Attack Simulation}

\renewenvironment{comment}{}{
\begin{figure}[t!]
\centering
	\includegraphics[width=0.4\textwidth]{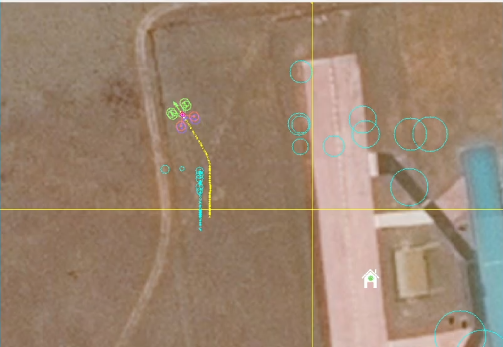}
	\caption{The drone is drifted away from its original path. The yellow line shows the drone's trajectory. The cyan circle denotes the obstacle position.}
	\label{pic21}
\end{figure}
}


We simulate \attack drone attacks using Ardupilot \cite{Ardupilot} and AirSim \cite{AirSim}. AirSim is an autonomous system simulator created by Microsoft, which is used to collect virtual-environment data in our simulation, while Ardupilot, a popular drone project, is used to simulate \attack on drones. 


\begin{figure}[t]
\centering
	\includegraphics[width=0.49\textwidth]{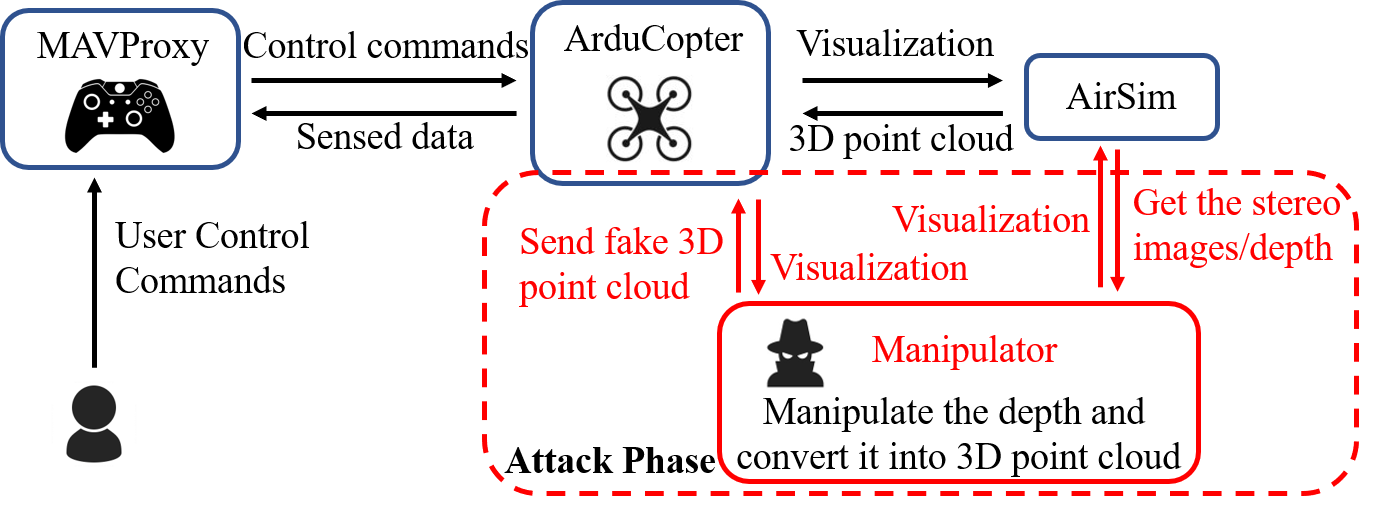}
	\caption{The simulation workflow of \attack with Ardupilot and AirSim.}
	\label{pic8}
\end{figure}

Fig. \ref{pic8} shows the workflow in the virtual environment. Ardupilot is used for simulating MAVProxy as the ground station, and ArduCopter as the drone. AirSim provides the sensor inputs to the ArduCopter. The user first sends commands to ArduCopter via MAVProxy. Next, ArduCopter sends its states to AirSim, which provides a simulated environment. After that, AirSim sends the sensor inputs back to ArduCopter, and the drone's OA system processes the received data and makes the flying decisions to avoid the obstacles.

\begin{figure}[b]
\centering
\begin{minipage}{0.5\textwidth}
  \centering
\subfloat[Flying trajectory on the zoom-in map in Ardupilot]{\includegraphics[trim={2mm 2mm 2mm 2mm},clip,width=0.47\textwidth]{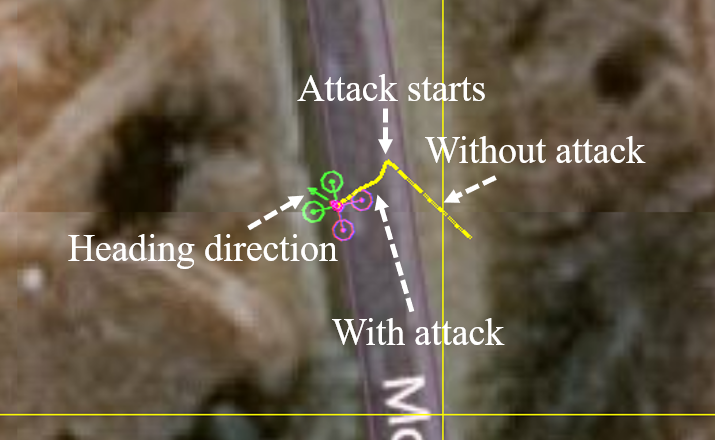}
\label{pic21a}}
\hspace{1mm}
\subfloat[Real flying environment in AirSim]{\includegraphics[width=0.46\textwidth]{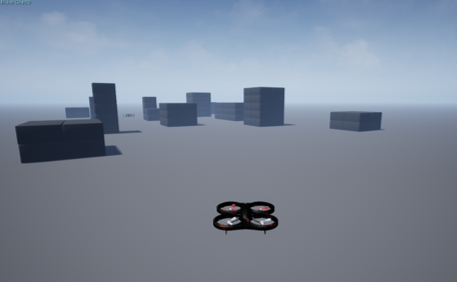}
\label{pic21b}}
\end{minipage}
\caption{The ArduCopter under attack  drifts away when there is no real obstacle near the drone.} \label{pic21}
\end{figure}

To attack the drone, we design a depth manipulator and embed it between ArduCopter and AirSim. By injecting different fake obstacle depths in a realistic scenario, we successfully demonstrate that our attack can achieve real-time drone control.
For example, to move the drone forward while ignoring the real obstacle in its path, we can generate a fake target that is apparently far away. The bright beams would overwhelm the sensors and make the actual barriers invisible. Conversely, we could also inject a seemingly close object to stop the drone. 
Remarkably, pushing the drone away from its original course is also possible if the attacker creates a fake object that floats at a constant distance from the drone by its side. Fig. \ref{pic21} shows such an example, where the attacker injects a fake depth to the front right position of the drone. The drone under attack drifts away from its heading direction to the left, even as there is no real obstacle present near the drone. \rev{We further demonstrate this attack in a real-world experiment in Section~\ref{endtoend}}. 

Other useful drone manipulations, such as drone body shaking and moving backward, are also feasible. 
For example, by merely injecting a fake object within the drone's OA distance, the drone will move backward. By manipulating the depth within its OA distance threshold, \attack could continue as the OA system attempts to steer the drone away. We consider two shaking patterns for shaking the drone: (1) front-to-back shaking and (2) left-to-right shaking. In the first case, we place the fake obstacle depth intermittently with a specific time interval, e.g., 0.5 seconds. When the drone detects the fake obstacle, it will retreat, only to revert course when the barrier suddenly disappears. Therefore, the attack forces the drone to go back and forth alternately, resulting in front-to-back body shaking. Similarly, we can also generate the fake depth on its front left and right alternately within a short time interval to shake the drone sideways. The demo video is available on the website. 

\minor{
\subsection{Attack Simulation on Stereo Depth Estimation Algorithms}\label{attack_depthAlg}

\textbf{Attacking Classic Algorithms.} We simulate the attack towards two classic stereo depth estimation algorithms, \emph{i.e.}, block matching (BM) \cite{StereoBM} and semi global block matching (SGBM) \cite{SGBM_opencv}. As shown in Fig. \ref{pic14}, we embed a patch on both images, which deceive both algorithms in generating unreliable depths. These classic non-AI-based depth estimation algorithms are still pervasive in real devices, due to their low computational complexity and short real-time delay~\cite{zhou2020review}. However, the state-of-the-art depth estimation algorithms are mostly driven by AI models, such as convolution neural network (CNN) and recurrent neural network (RNN)~\cite{laga2020survey}. These algorithms leverage deep neural network (DNN) to learn representations from image data to extract the depth information. 

\begin{figure}[]
\centering
\begin{minipage}{0.5\textwidth}
  \centering
\subfloat[Attack BM]{\includegraphics[width=0.95\textwidth]{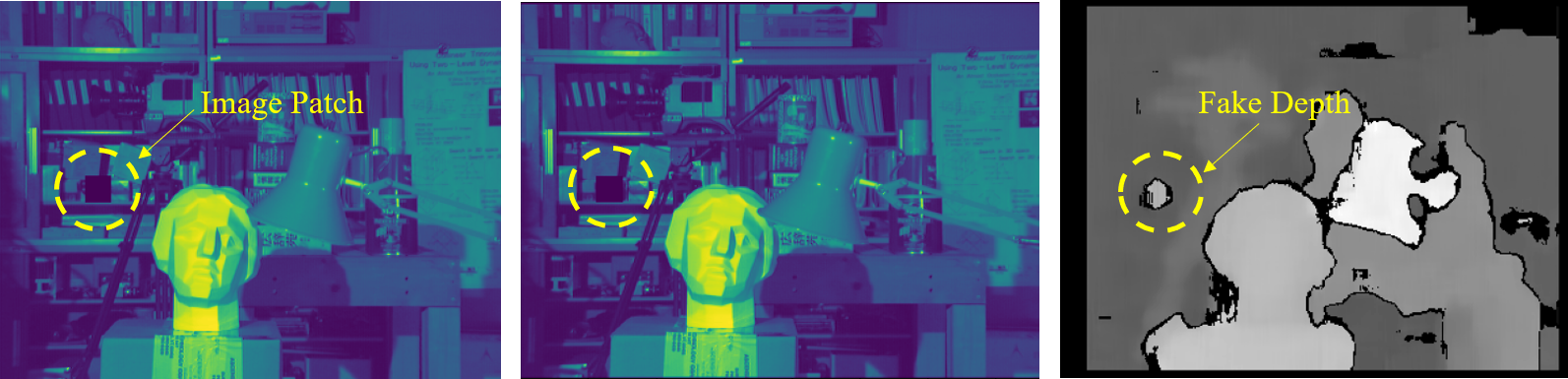}
\label{pic14a}}

\subfloat[Attack SGBM]{\includegraphics[width=0.95\textwidth]{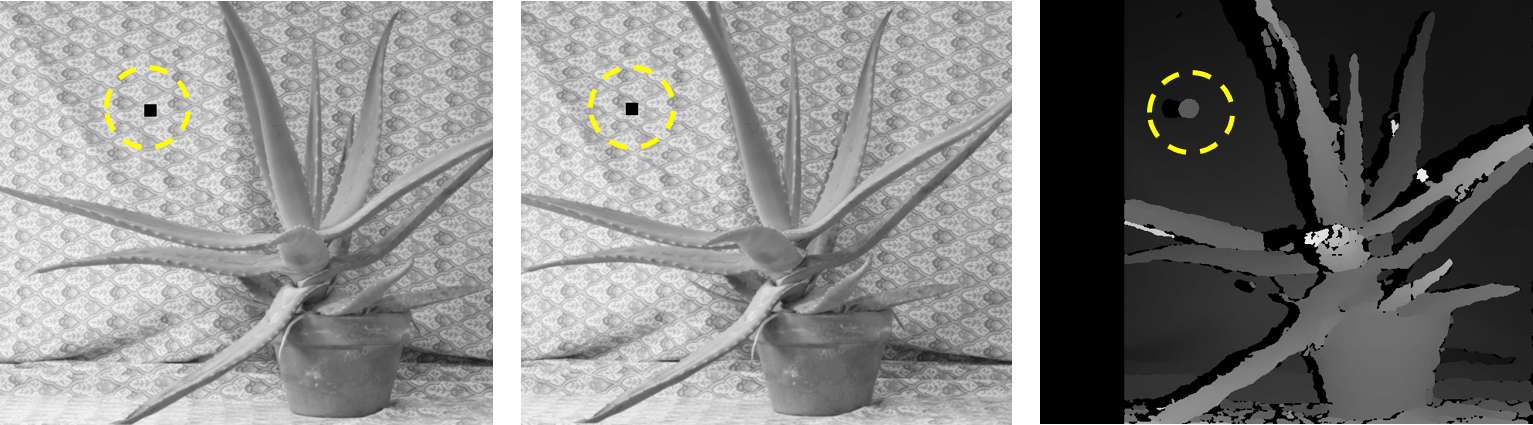}
\label{pic14b}}
\end{minipage}
\caption{An image patch is attached to left and right images in both BM and SGBM depth estimation algorithms, and the result confirms that our attack can compromise these depth estimation algorithms.} 
\label{pic14}
\end{figure}

\textbf{Attacking AI-based Algorithms.} To verify the generality of \attack, we test the attack on three state-of-the-art AI-based stereo depth estimation algorithms, \emph{i.e.}, DispNet~\cite{mayer2016large}, PSMNet~\cite{chang2018pyramid}, and AANet~\cite{tay2019aanet}. DispNet is an end-to-end trainable framework for depth estimation, where a correlation layer is used to measure the similarity of left and right image features. PSMNet takes a different approach by directly concatenating left and right features, and then 3D convolutions are used to aggregate the costs to achieve higher accuracy. AANet uses a cost aggregation method based on sparse points 
in conjunction with neural network layers to achieve a faster inference speed while maintaining comparable accuracy.

\begin{figure}[t]
\centering
\begin{minipage}{0.5\textwidth}
  \centering
\subfloat[X-shape beams attack]{\includegraphics[width=0.95\textwidth]{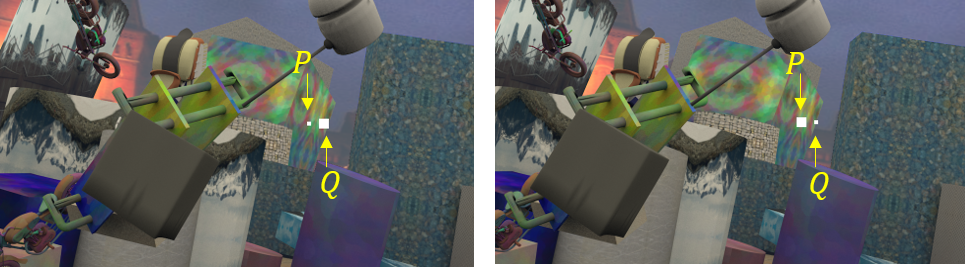}
\label{pic30}}

\subfloat[Trapezoid-shape beams attack]{\includegraphics[width=0.95\textwidth]{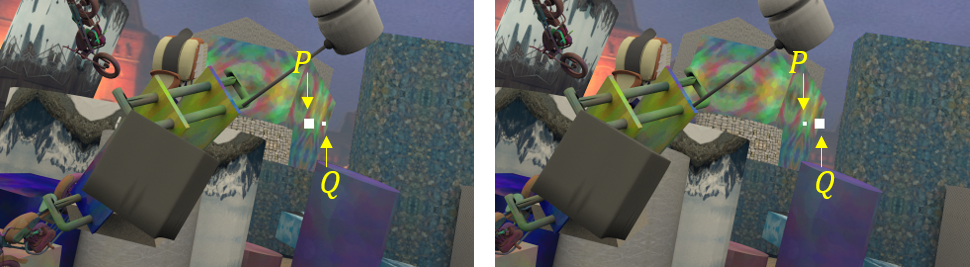}
\label{pic31}}
\end{minipage}
\caption{\rev{Two image patches, $P$ and $Q$, are attached to left and right benign images.}} 
\label{fig:attack_patch}
\end{figure}

Fig. \ref{fig:attack_patch} shows examples of X-shape and trapezoid-shape beams attacks. From Figs. \ref{pic30} and \ref{pic31}, we can see two small image patches are embedded in the stereo image, corresponding to the two light beams (\emph{i.e.}, $P$ and $Q$).
We take this adversarial stereo image pair as the input to the three algorithms. The corresponding outputs of the three algorithms are shown in the Fig. \ref{fig:attack_depth_map}. 
We can see that all three algorithms can be deceived by X-shape beams attack, since a near-distance fake obstacle is generated as expected. Regarding the trapezoid-shape beams attack, we expect to see a far-distance fake depth in the image according to our mathematical analysis. However, since a far-distance fake obstacle is blended into the background in the depth map, it can hardly be observed. 
Note that, the fake depth value depends on the position of the injected patches, e.g., by separating $P$ and $Q$ away from each other in the stereo image pair, the fake depth value grows.

\begin{figure}[t]
\centering
\begin{minipage}{0.5\textwidth}
  \centering
\subfloat[DispNet]{\includegraphics[width=0.33\textwidth]{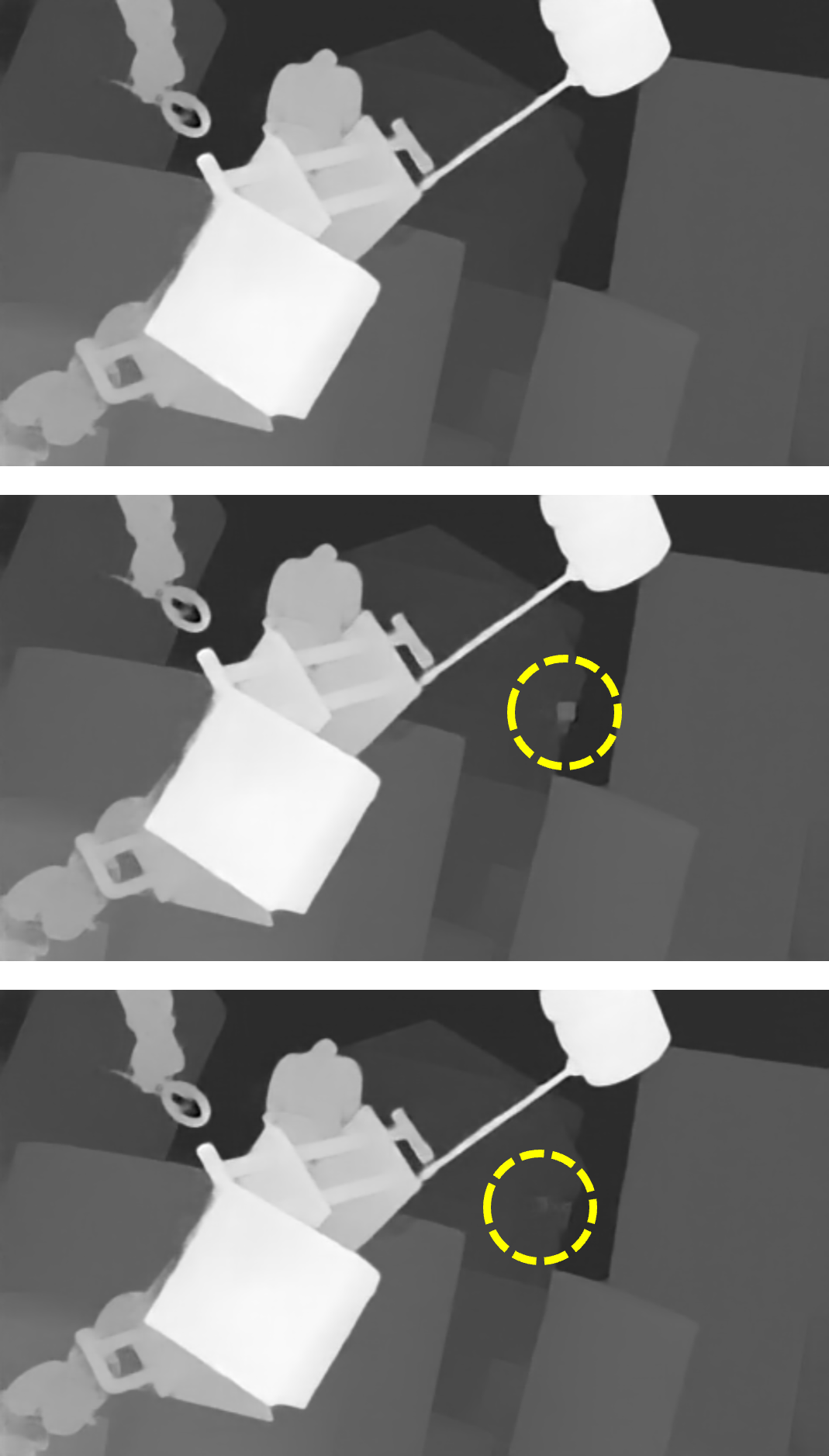}
\label{pic32}}
\subfloat[PSMNet]{\includegraphics[width=0.33\textwidth]{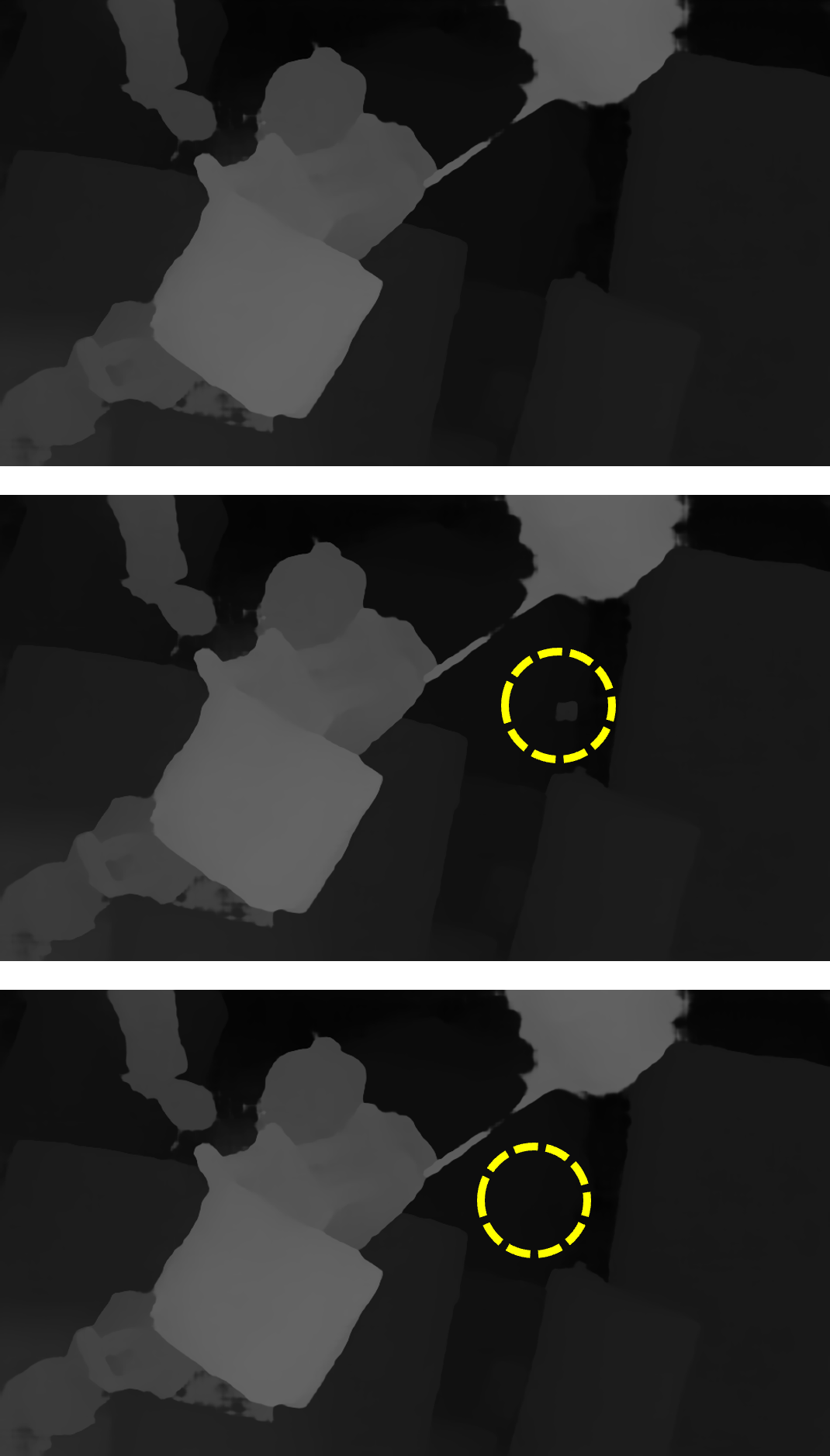}
\label{pic33}}
\subfloat[AANet]{\includegraphics[width=0.33\textwidth]{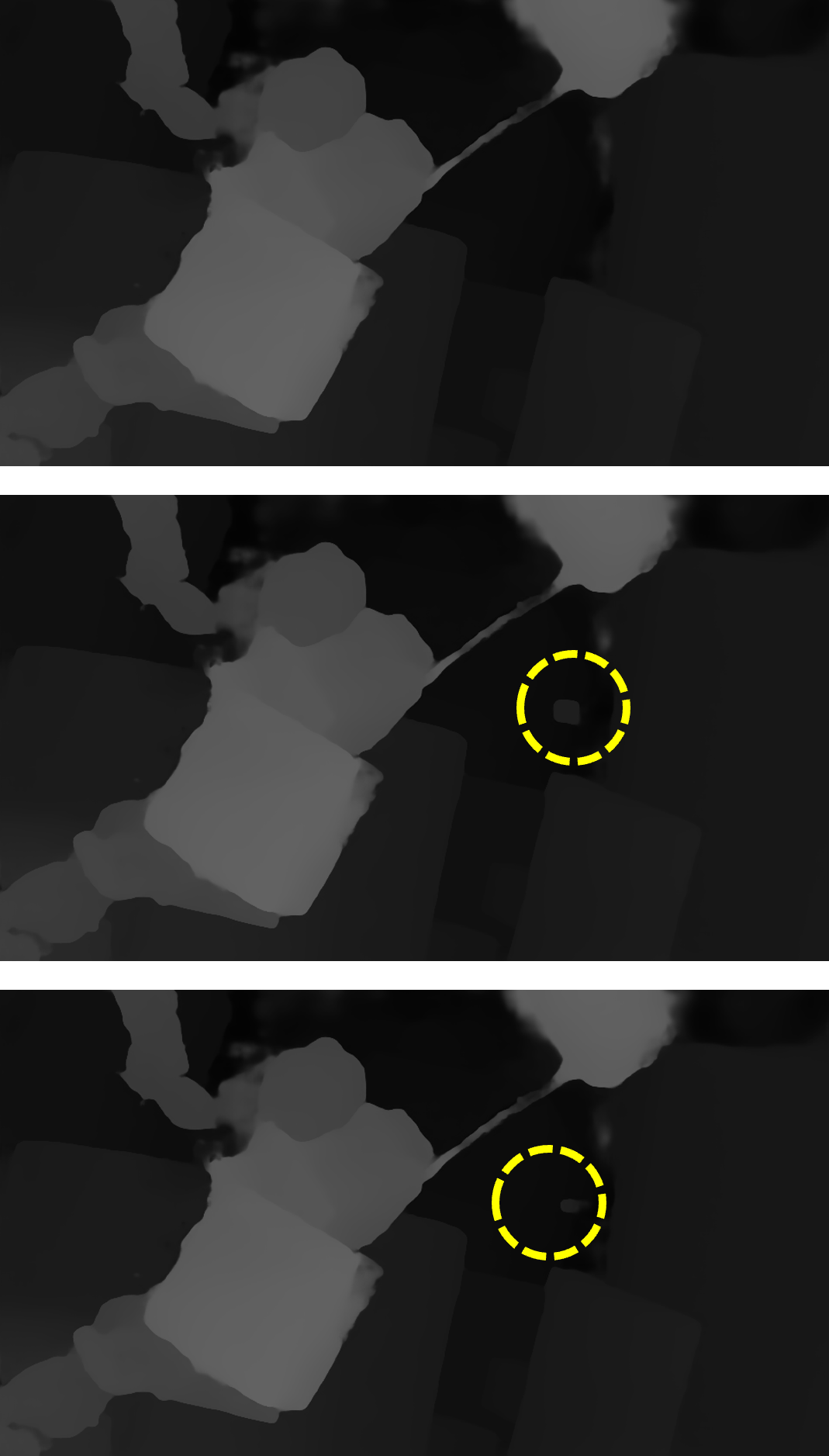}
\label{pic34}}
\end{minipage}
\caption{\rev{Depth maps from DispNet, PSMNet, and AANet. Each column, from up to down, represents benign depth map, depth map from X-shape beams attack, and depth map from trapezoid-shape beams attack. The fake depths are circled.}} 
\label{fig:attack_depth_map}
\end{figure}

\textbf{Verifying the Attack.} We collect the adversarial stereo images from the real-world attack and use them as input to these three algorithms. Specifically, we use the stereo image pairs of X-shape beams attack and trapezoid-shape orbs attack in Figs.~\ref{pic4a},~\ref{pic4d} as input. Their corresponding fake depth maps from the three algorithms are shown in Fig.~\ref{fig:real_attack_depth_map}. It can be seen that except the orbs attack on AANet, all the attacks successfully inject fake depth information. For the orbs attack on AANet, the orbs disappear from the depth map, which may have been smoothed out by the AANet algorithm.   

\renewenvironment{comment}{}{
\begin{figure}[H]
\centering
\begin{minipage}{0.5\textwidth}
  \centering
\subfloat[Trapezoid-shape orbs attack]{\includegraphics[width=0.95\textwidth]{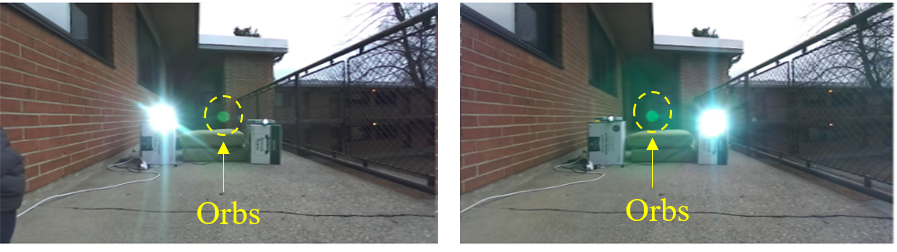}
\label{pic35}}

\subfloat[X-shape beams attack]{\includegraphics[width=0.95\textwidth]{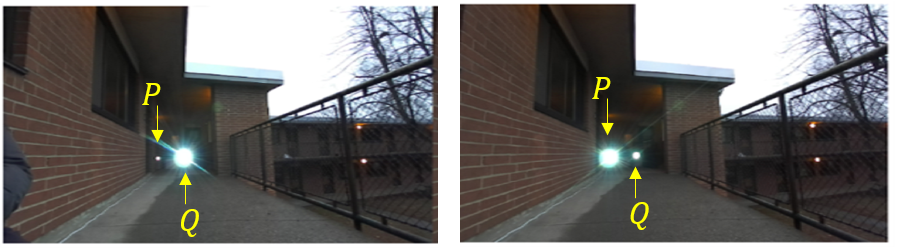}
\label{pic36}}
\end{minipage}
\caption{\rev{Attack image pairs from ZED in real life attack scenarios.}} 
\label{fig:real_attack_data}
\end{figure}
}

\begin{figure}[t]
\centering
\begin{minipage}{0.5\textwidth}
  \centering
\subfloat[DispNet]{\includegraphics[width=0.33\textwidth]{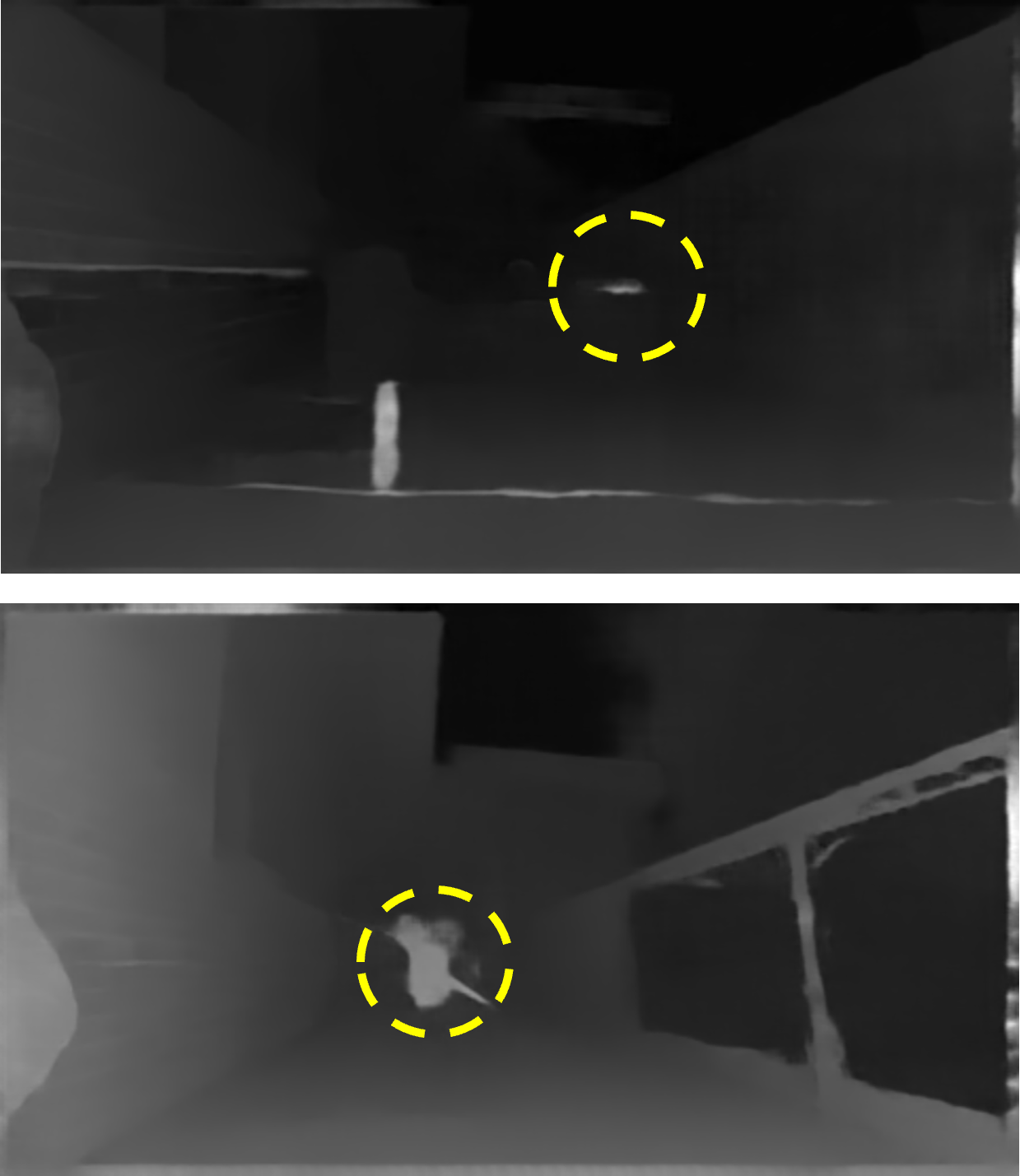}
\label{pic37}}
\subfloat[PSMNet]{\includegraphics[width=0.33\textwidth]{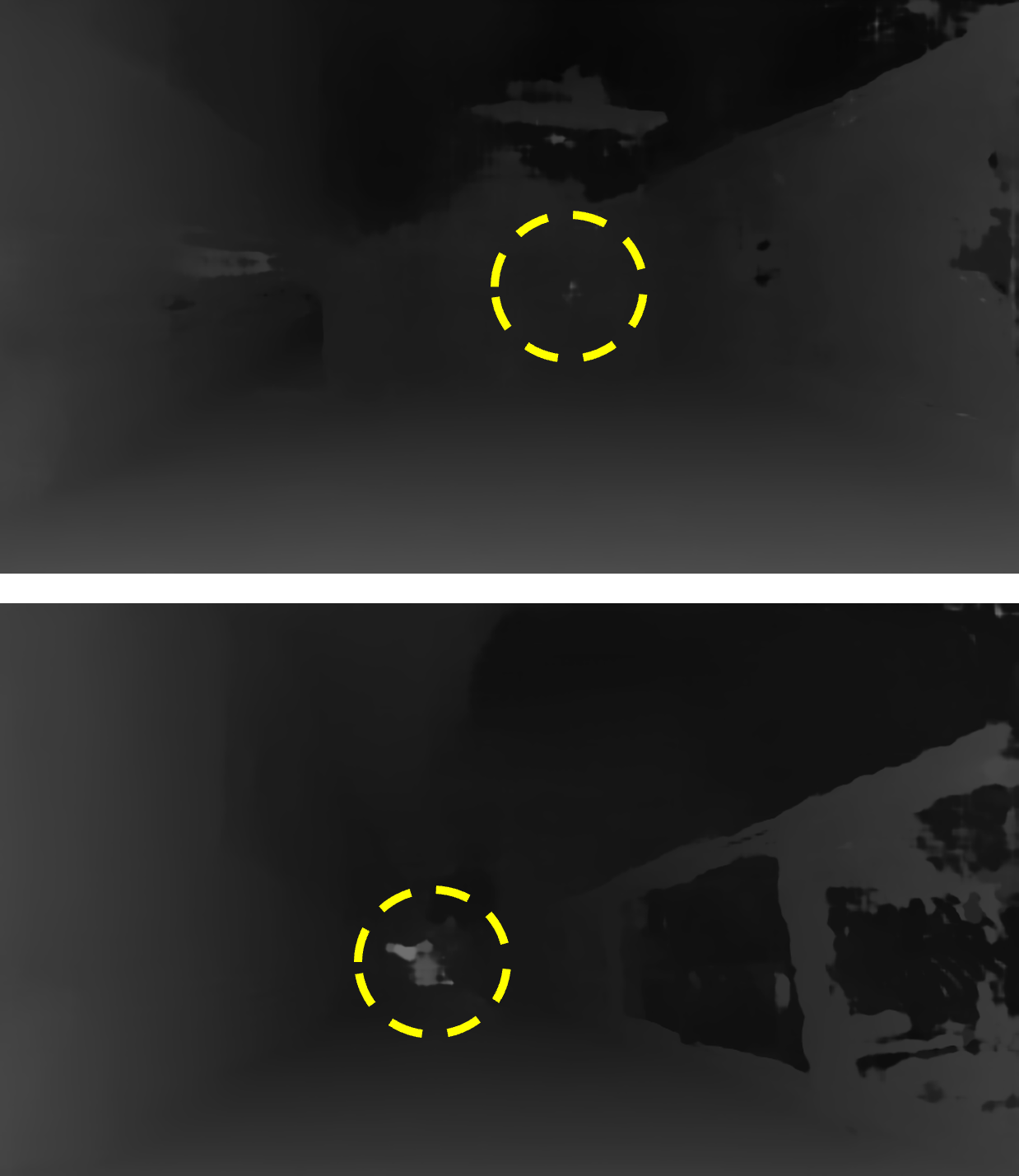}
\label{pic38}}
\subfloat[AANet]{\includegraphics[width=0.33\textwidth]{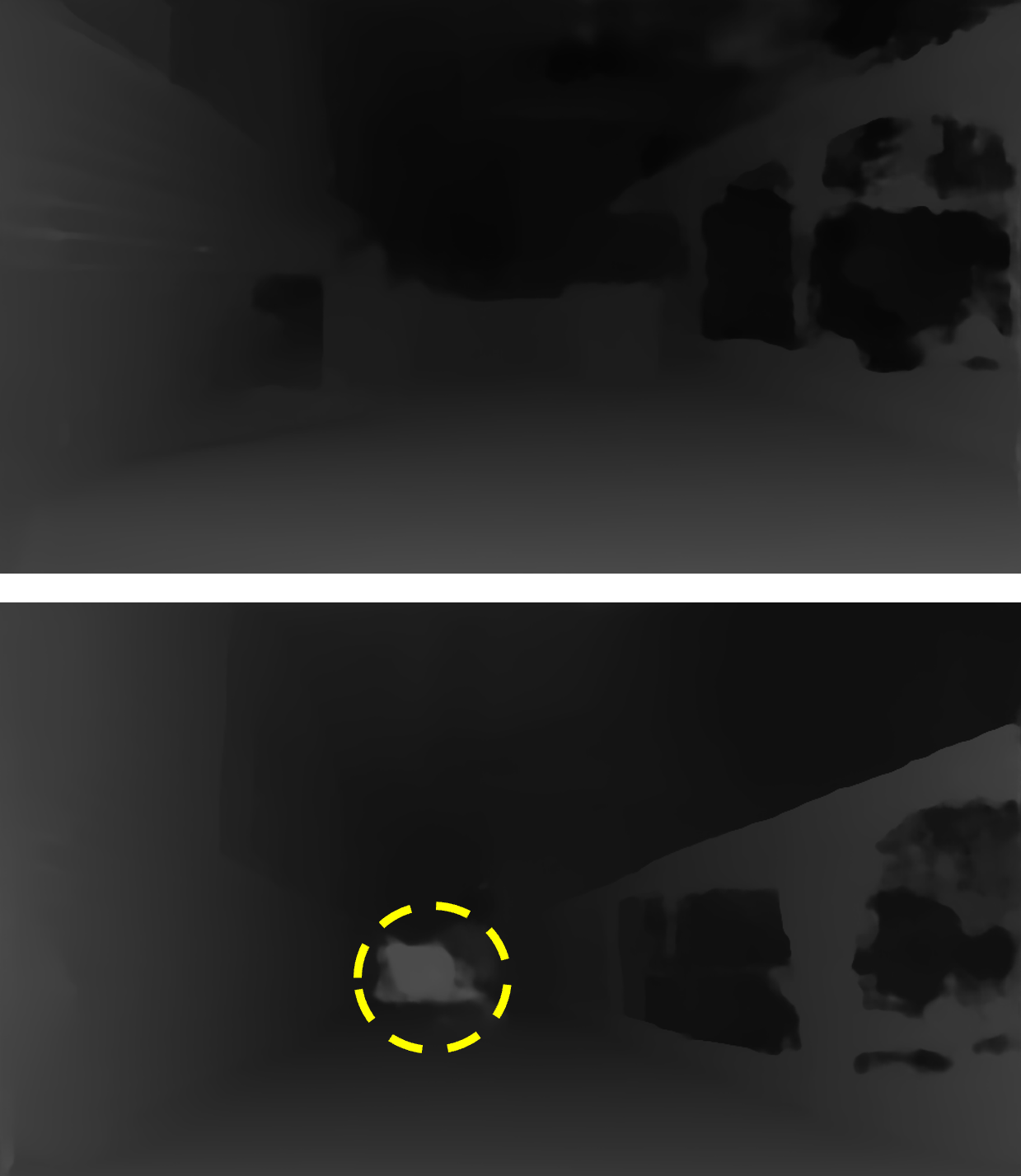}
\label{pic39}}
\end{minipage}
\caption{\rev{Depth maps from DispNet, PSMNet, and AANet. For each column, the upper depth map is from X-shape beams attack, and the lower one is from trapezoid-shape orbs attack. The fake depths are circled.}} 
\label{fig:real_attack_depth_map}
\end{figure}

}

\section{Evaluation}\label{evaluation}

In this section, we evaluate \attack on the depth estimation-based OA systems used in autonomous systems. 
Specifically, we showcase proof-of-concept \attack 
on a commercial drone, DJI Phantom 4 Pro V2, and two stereo cameras, ZED and Intel RealSense D415. For simplicity, we refer DJI Phantom 4 Pro V2 as the DJI drone, ZED stereo camera as ZED, and Intel RealSense D415 stereo camera as RealSense. We select  DJI drone due to its high popularity 
and state-of-the-art stereo vision based OA systems\cite{DJI_Phantom_4}. ZED and RealSense are chosen since they are specially designed for autonomous robotic systems, both of which use the cutting-edge AI-based algorithms to compute the depth~\cite{ZED, Intel}. 

The experiments aim to measure (1) the range within which the fake obstacle can be generated, (2) the range of attack distance, and (3) the range of attack angle within which we can successfully launch the attacks.

\begin{figure}[t]
\centering
	\includegraphics[width=0.46\textwidth]{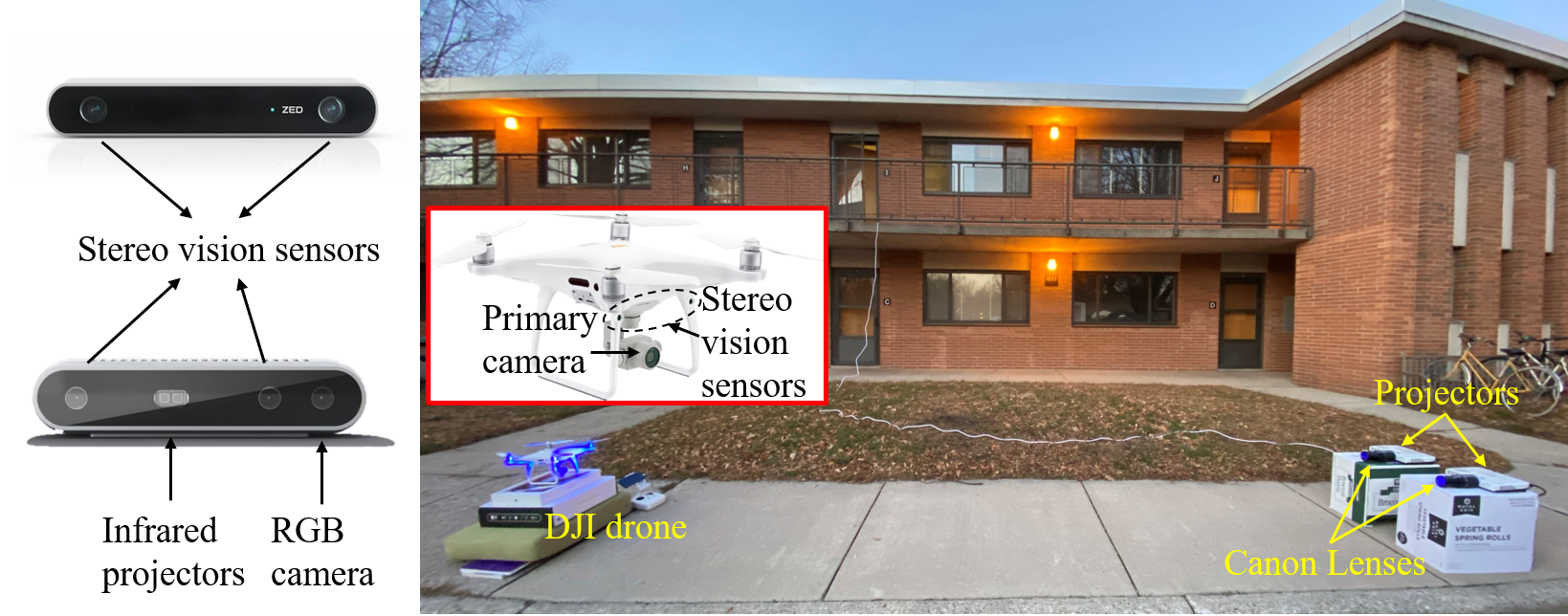}
	\caption{Outdoor attack experimental setup: the DJI drone is on the left and two Epson PowerLite 1771W projectors \cite{Epson} combined with two Canon EF $75-300 mm$ zoom lenses are on the right. We also conduct the experiments on ZED and RealSense in the experiments. All three attack devices have a pair of stereo vision sensors. The DJI drone has a primary camera at the bottom and RealSense has extra infrared projectors and an RGB camera.}
	\label{pic9}
\end{figure}

\subsection{Experimental Setup}

The evaluation setup is shown in Fig. \ref{pic9}, where we have the DJI drone on the left and two projectors combined with two zoom lenses on the right in an outdoor environment. The DJI drone could be switched into ZED or RealSense according to the setup of different experiments. The \emph{throwing ratio} is defined as the ratio between the projection distance and the size of the projection screen. If an attacker aims to perform long-distance attacks, he/she would need the projector to have a larger throwing ratio to concentrate the light beams. 

The size of the lenses on the stereo camera is tiny, e.g., the area of the lens on ZED is $1.3 \times 10^{-4} m^2$. The throwing ratio of our projector is also very small, i.e., 1.04 – 1.26, which means that the projection screen area is at most 0.62 $m^2$ at 1$m$. 
Comparing $1.3 \times 10^{-4} m^2$ with 0.62 $m^2$, we can see that less than 1\% of the projection light can be injected into the lens. Therefore, to further extend the attack distance, we integrate each projector with a Canon zoom lens (\emph{i.e.}, Canon EF 75-300mm) \cite{man2020ghostimage}. 
In our experiments, the maximum throwing ratio is increased to 30, when the focal length is adjusted to 300 $mm$, which implies that $2.5 \times 10^{-3} m^2$ projection screen area can be achieved at 1$m$. Thus, around 12\% of the projection light can arrive at the lens,
making the long-range attack feasible.

Since we have no access to the DJI drone’s sensor data, we use the DJI drone to read the depth in meters and use the ZED to verify the left and right images, depth map, and 3D point cloud. 
For RealSense, as it does not give us access to its left and right images, we will only check the depth map to verify the success of our attack. For the experiments on the DJI drone, we define the attack as successful if the generated fake depth is less than 6$m$, which is the threshold value to trigger actions of the OA system. For the experiments on ZED and RealSense, we record it as a success as long as a near-distance fake depth can be seen in the depth map. We perform each attack pattern 3 times in every experiment.

During the experiments, the distance between the two projectors is fixed as $1 m$. The environmental ambient light levels are $4000 lux$ and $0 lux$ for day and night, respectively. All the experiments are conducted outdoor. Unless otherwise specified, these parameters are the default for all experiments.


\subsection{Fake Depth Range}

Table \ref{tab2} summarizes the range of fake obstacle depth that the attacks can achieve on the DJI drone during the daytime and at night. The default value of the projection light illuminance is $1.6 \times 10^4$ $lux$ without connecting to any source. We perform each attack pattern with different attack distances and record all possible fake depths it generates.

\begin{table}[t]
\centering
\rev{
\caption{\rev{Fake depth range on the DJI drone at night and during the daytime with different attack patterns. Expected fake depth is derived from the mathematical model. ``None" means no fake depth can be successfully injected.} }
\label{tab2}
\resizebox{0.48\textwidth}{!}{%
\begin{tabular}{|l|l|l|l|l|l|l|l|l|}
\hline
\multirow{2}{*}{\begin{tabular}[c]{@{}l@{}}\textbf{Attack} \\ \textbf{Distance ($m$)}\end{tabular}} &
  \multicolumn{2}{l|}{\begin{tabular}[c]{@{}l@{}}\textbf{Expected Fake} \\ \textbf{Depth ($m$)}\end{tabular}} &
  \multicolumn{3}{l|}{\textbf{Fake Depth Range at Night ($m$)}} &
  \multicolumn{3}{l|}{\textbf{Fake Depth Range in the Day ($m$)}} \\ \cline{2-9} 
   & \textbf{X}      & \textbf{Trapezoid} & \textbf{X}       & \textbf{Trapezoid} & \textbf{Triangle}  & \textbf{X}         & \textbf{Trapezoid} & \textbf{Triangle}   \\ \hline
1  & 0.5    & 0.5       & 0.5     & 0.5       & None      & 1.5       & 1         & None       \\ \hline
2  & 0.5    & 0.5       & 0.5     & 0.5       & None      & 0.5 - 1   & 8         & None       \\ \hline
3  & 0.5    & 0.5       & 0.5     & 0.5       & 0.5 - 16  & 1 - 1.5   & 3         & None       \\ \hline
4  & 0.5, 1 & 1         & 0.5 - 1 & 10.5      & 0.5 - 16  & 0.5 - 1.5 & 10 - 11   & 0.5 - 11   \\ \hline
5  & 0.5    & 1         & 0.5     & 11        & 0.5 - 16  & 0.5 - 2   & 5 - 11    & 1 - 5.5    \\ \hline
6  & 0.5    & 1         & 0.5 - 1 & 12.5      & 10.5 - 16 & 1 - 2     & 5 - 11    & 1 - 11     \\ \hline
7  & 0.5    & 1         & 1       & 12 -16    & 6.5 - 16  & None      & None      & 0.5 - 14.5 \\ \hline
8  & 1      & 1         & 0.5-1   & 12-16     & 6.5 - 16  & None      & None      & 1 - 14     \\ \hline
9  & 1      & 1         & None    & None      & 1 - 16    & None      & None      & None       \\ \hline
10 & 1      & 1.5       & None    & None      & 1.5 -10.5 & None      & None      & None       \\ \hline
11 & 1      & 1.5       & None    & None      & 1 - 16    & None      & None      & None       \\ \hline
12 & 1.5    & 1.5       & None    & None      & 1.5 - 16  & None      & None      & None       \\ \hline
13 & 1.5    & 2         & None    & None      & 1.5 - 16  & None      & None      & None       \\ \hline
14 & 1.5    & 2         & None    & None      & 10.5 - 14 & None      & None      & None       \\ \hline
15 & 1.5    & 2         & None    & None      & 16        & None      & None      & None       \\ \hline
16 & 1.5    & 2         & None    & None      & None      & None      & None      & None       \\ \hline
\end{tabular}%
}
}
\end{table}

The results show that \attack can achieve maximally $13 m$ attack distance when the fake depth is under 6\emph{m} (i.e., a successful attack). Our attacks can achieve up to $15 m$ in distance, in which case the depth of a real near-distance object is converted into the depth of a far-away fake obstacle. Using all three attack patterns, our attack can generate various fake depths with the attack distance ranging from $3-8 m$ at night and $4-6 m$ in the day. However, there are a few cases when only partial attack patterns work. Specifically, when the attack distance is $1-2 m$, the triangle-shape attack fails in both ambient light conditions. The reason is that in these cases the width of the projection screen is smaller than the baseline, and no projection light can enter the stereo camera.
When the attack distance is $3 m$, the triangle-shape attack works at night but fails in the day due to the strong ambient light. 
In fact, during the daytime, only the marginal projection light can enter the stereo camera, resulting in an injected light intensity that is too weak to deceive the depth perception.
Also, when the attack distance is more than $8 m$ at night and $6 m$ during the daytime, both trapezoid- and X-shape attacks fail. The reason is that the divergent light beams traversing a long distance significantly weaken the injected light intensity. 

\minor{It is worth noting that modern cameras  are usually equipped with the auto exposure (AE) control mechanisms \cite{lee2005introduction}, which automatically balance the brightness of the captured image. The exposure increases if the overall brightness turns dark, and vice versa. 
Our results show that the orbs attacks usually fail during the day, while the beams attacks succeed. This phenomenon is likely caused by AE, when the brightness of the injected beams induces a drop of lightness in the image background. As a result, the orbs become less visible. 
}

Moreover, in order to launch the trapezoid- and X-shape attacks, the attacker should avoid lights overlapping at the drone side. To achieve that, the attacker can only use marginal light to launch these two attack patterns. As a result, the injected light intensity becomes too weak to attack effectively. However, overlapping is not an issue for the triangle-shape attack. 
That is why the triangle-shape attack can achieve the longest attack distance. 
\emph{In summary, the range of fake depth is $0.5-16 m$. This range covers all the possible depths that can be sensed by DJI drone’s OA system, which makes real-time drone control possible.}

\begin{figure}[t]
\begin{minipage}{0.48\linewidth}
\centering
	\subfloat[Attack range on the DJI drone.]{\includegraphics[trim=3mm 1mm 1mm 6mm, clip,width=\textwidth]{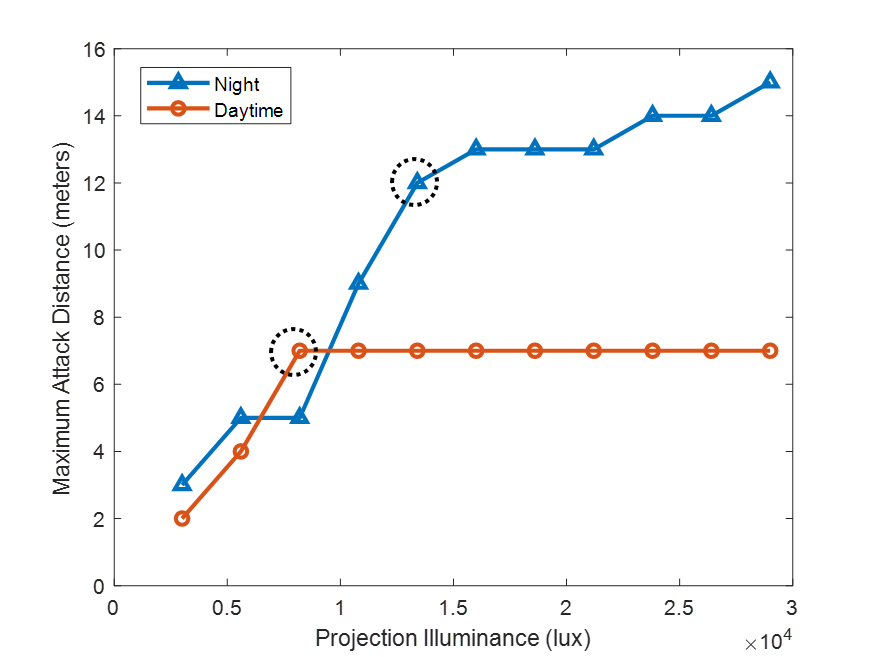}
	\label{pic10}
	\hspace{8pt}}
    \subfloat[Attack range on ZED.]{\includegraphics[width=\textwidth]{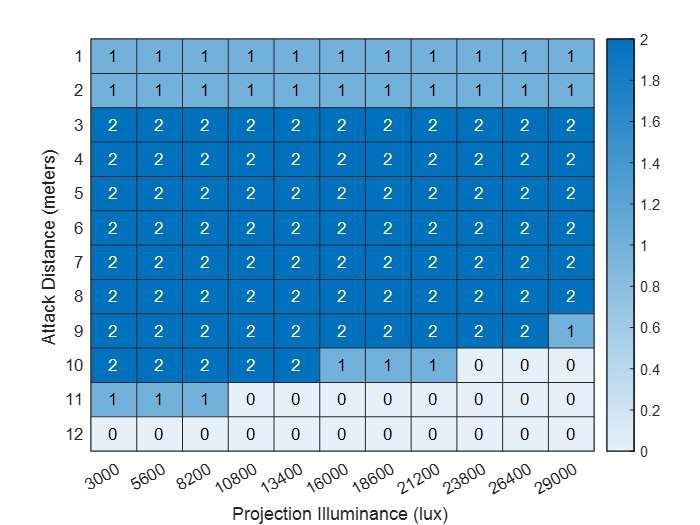}
	\label{pic11}}
\end{minipage}
\caption{Attack range w.r.t. projection illuminances.}
\label{fig:attack_range}
\end{figure}

\subsection{Range of Attack Distance}

The range of attack distance is the key evaluation criterion in our attack, since \emph{no prior work has ever achieved a long-range drone sensor attack}. With the projector's default light illuminance, we can achieve up to $13 m$ attack range at night and $7 m$ attack range during the daytime. Further, we explore the impact of the projection light intensities on the range of attack distance on both ZED and DJI drone.

\begin{figure*}[t!]
\centering
\subfloat[Illustration of three attack angles]{\includegraphics[width=0.24\textwidth]{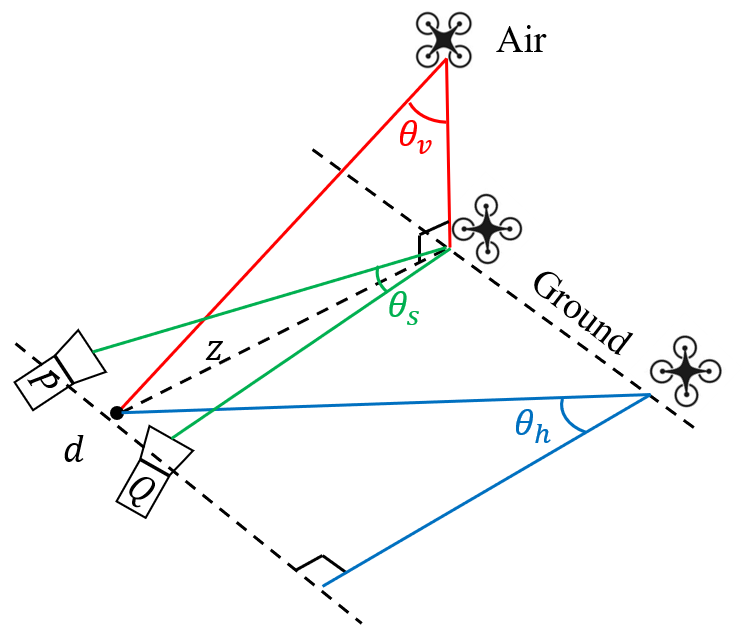}
\label{pic40}}
\subfloat[Horizontal attack angle $\theta_h$]{\includegraphics[width=0.24\textwidth]{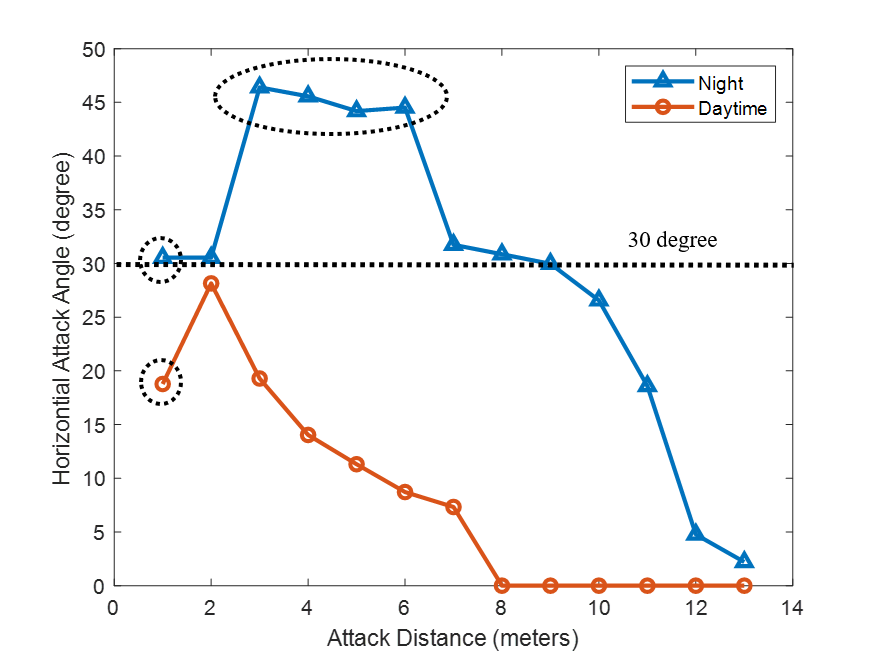}
\label{pic24}}
\subfloat[Vertical attack angle $\theta_v$]{\includegraphics[width=0.24\textwidth]{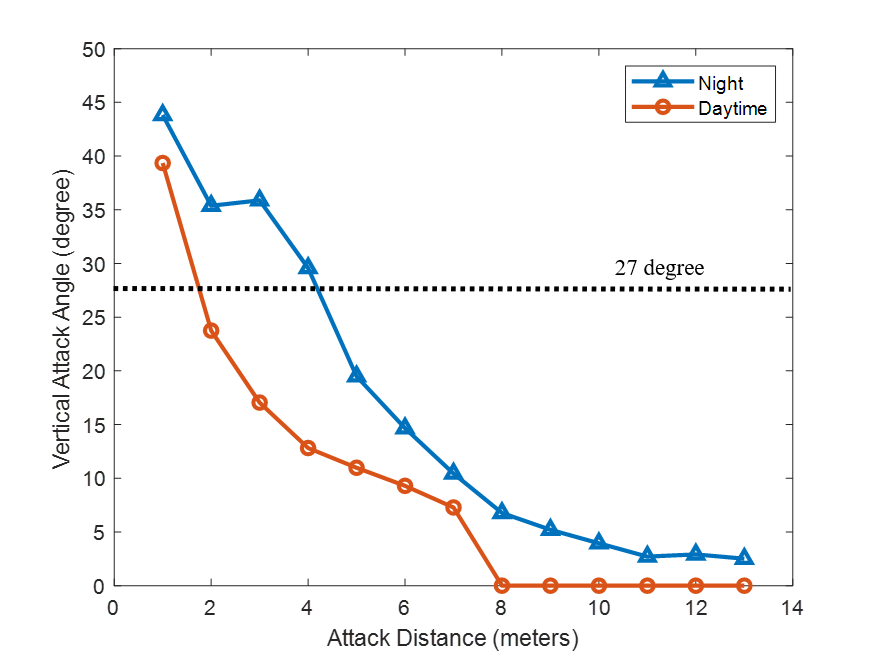}
\label{pic25}}
\subfloat[Spinning attack angle $\theta_s$]{\includegraphics[width=0.24\textwidth]{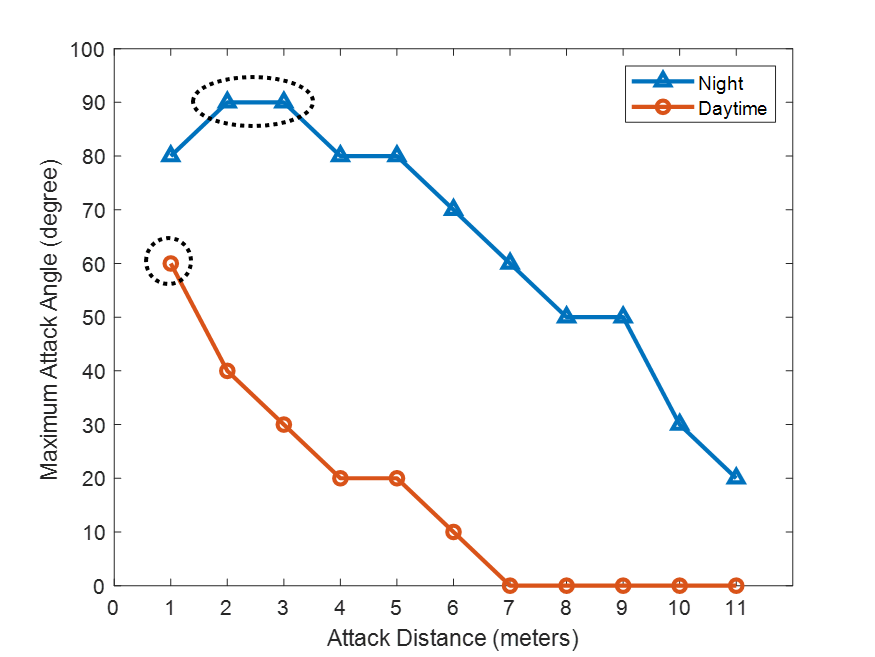}
\label{pic12}}
\caption{\rev{The maximum attack angle with varying achievable attack distances during the day and at night.}}
\label{fig:range}
\end{figure*}

\textbf{Attack Range Results from Drone.}
We repeat the experiments with different projector intensities on the DJI drone, and record the longest distance where our attack is successful as the corresponding attack range. 

\rev{Fig. \ref{pic10} presents the attack distance range on DJI drone with various projection intensities. Each point in the figure refers to the longest attack distance at which the fake depth can be observed. We determine an attack as successful when the fake depth is observed by the controller.} The results show that at night our attack can achieve up to $15 m$ with the highest projection intensity, and up to $3 m$ with the lowest projection intensity. The attack range increases dramatically with the increasing of the projection illuminance below $1.4 \times 10^4 lux$ and grows smoothly afterwards. The reason behind it is that the projection illuminance has a more dominant impact than the attack distance at the beginning since the light is very concentrated; whereas, the light beam becomes more divergent beyond $12 m$, 
which limits the injected light intensity even with a higher projection illuminance. On the other hand, our attack can be up to $7 m$ during the day. The attack range increases proportionally with the increasing of the projection intensity at the beginning. However, when the luminosity reaches $8.4 \times 10^3$ $lux$, the attack range value plateaus and remains the same ever since. It is because as the distance increases, the light source becomes less predominant in the image. \minor{In addition, due to the AE control under the strong ambient light, the camera could view the background environment more clearly.} Thus, false stereo correspondence matching is avoided.

Our results indicate that: \emph{(1) as the perpendicular distance between the DJI drone and the projector increases, a stronger projection intensity is required to generate the fake obstacle depth; (2) launching successful \attack at night is easier than during the daytime due to the influence of the ambient light; (3) during the daytime, even with a stronger projection intensity, it is very difficult to achieve a larger attack distance because the projected light source becomes less predominant and the environment becomes clearer in the image.}

\textbf{Visualization Results from Stereo Camera.}
For the attack on ZED, we visualize the image depth map and 3D point cloud to evaluate the range of attack distance and explore the relationship between the beams attack and orbs attack. 

We conduct the experiments only in the daytime to better observe the depth. 
By repeating the experiments with different projection intensities, we record all attack results for different attack distances using shades of blue as shown in Fig. \ref{pic11}. In the figure, ``2" indicates the case when fake depths from both the beams and orbs attack are observed; ``1" indicates that only fake depth from the orbs attack is observed; ``0" indicates no fake depth is observed. It can be seen that when the attack distance is $1-2 m$, only the orbs attack works. It is because the disparity of these two projectors is too large to be matched as the same object, which has been discussed in Section \ref{math_beams}. When the distance is from $3-8 m$, both the attacks can be observed in the depth map and 3D point cloud. However, when the distance increases to $9-11 m$, the fake depth can still be observed from the orbs attack with stronger light intensities, but not from the beams attack. This can be attributed to the weak light intensity which is insufficient for beams attack to succeed, but the lens flare effect is unaffected. Note that both attacks are invisible with a weaker projection intensity because of the low luminosity. When the distance goes beyond $12 m$, the light is too weak to execute any successful attack due to the more divergent light beam. We observed that within the successful range, the orbs attack usually outperforms beams attack within the short attack range, whereas beams attack becomes more evident as the attack range increases.

Our results indicate that: \emph{(1) merging the beams attack with orbs attack helps increase the attack range; 
(2) the orbs attack is more resistant to the weak projection intensity than the beams attack; (3) \attack can achieve up to $11 m$ in the day with a strong projection intensity on ZED.}

\renewenvironment{comment}{}{

\begin{figure}[t]
\centering
\begin{minipage}{0.5\textwidth}
  \centering
\subfloat[Attacking the backward vision sensors during the day]{\includegraphics[width=0.6\textwidth]{figures/Picture19a.png}
\label{pic19a}}

\subfloat[Attacking the forward vision sensors at night]{\includegraphics[width=0.6\textwidth]{figures/Picture19b.png}
\label{pic19b}}
\end{minipage}

\caption{} 
\label{pic19}
\end{figure}
}

\rev{

\subsection{Relative Positions of Attacker and Drone}\label{exp:position}

In this section, we evaluate the attack performance with respect to the relative positions of the attacker and drone. Specifically, 
we define three types of attack angles, including \emph{horizontal attack angle}, \emph{vertical attack angle} and \emph{spinning attack angle}. As shown in Fig. \ref{pic40}, we define the horizontal and vertical angle as the included angle between the center point of the two attack projectors and the attack target, denoted as $\theta_h$ and $\theta_v$ respectively. Spinning attack angle is defined as the included angle between the two projectors and the attack target ($\theta_s$) at the ground plane. We perform the experiments on DJI drone to evaluate the impact of attack angles. 

In all the following experiments, we fix the attack range and change the horizontal/vertical/spinning attack angles. Then, we record the maximum attack angle to launch a successful attack, with respect to the varying attack distances $z$ between the attacker and drone in Fig. \ref{pic40}.


\textbf{Horizontal Attack Angle $\theta_h$.}
The horizontal field of view (FOV) of the DJI drone is 60\textdegree \cite{DJI_Phantom_4}, i.e., when the $\theta_h$ is more than 30\textdegree, both projectors are out of the sensors' view. Fig. \ref{pic24} shows maximum $\theta_h$ with respect to different attack distances. 

During the day time, $\theta_h$ is 18\textdegree$ $ at $1 m$. The largest attack angle (29\textdegree) can be achieved when the attack distance is $2m$. However, $\theta_h$ decreases beyond $2m$, mainly due to the increasing straight-line distance between the camera and the projector. A longer distance results in weaker injected light, which in turn leads to a smaller attack angle. Note that since the distance between the two projectors is $1m$, when the attack angle is 18\textdegree$ $ at $1m$, one projector is already out of the vision sensors' view while the other one is still in the view. Even when the light source is out-of-view, several orbs can still be generated due to the out-of-view lens flare effect  (see Appendix \ref{out_of_view}), resulting in a successful orbs attack. 
Moreover, the light beam is more concentrated at 1$m$, making it harder to inject light into the stereo camera with a wider attack angle. 

At night, we can see the overall attack performance is better than that  during the day because of the absence of the ambient light. 
Most of the attack angles are around 30\textdegree or below. Beyond $7m$, $\theta_h$ decreases dramatically due to the decrease of the injected light intensity. However, when the attack distance is $3-6m$, $\theta_h$ reaches around 45\textdegree, the largest $\theta_h$ at night, in which case both projectors are out of the sensors' view. This attack is thus the result of out-of-view lens flare effect. 

\textbf{Vertical Attack Angle $\theta_v$.}
The vertical FOV of the DJI drone is +27/-27\textdegree \cite{DJI_Phantom_4}, \emph{i.e.}, when $\theta_v$ is more than 27\textdegree, both projectors are out of the vision sensors' view. Fig. \ref{pic25} shows maximum $\theta_v$ with varying  attack distances. The range of $\theta_v$ during the day and night have very similar trend while the performance in the day outperforms that at night. It can be observed that when the attack distance is at $1m$ in the daytime and $1-4m$ at night, $\theta_v$ is larger than the vertical FOV, which is mainly caused by the out-of-view lens flare effect. The maximum $\theta_v$ is around 40\textdegree$ $ during the day and 45\textdegree$ $ at night. With the increasing attack distance, $\theta_v$ in both scenarios shrinks due to the drop of injected light intensity. 

\textbf{Spinning Attack Angle $\theta_s$.}
Fig. \ref{pic12} shows the maximum $\theta_s$ with varying attack distance.  
During the day, we can see that $\theta_s$ is 60\textdegree$ $ at $1 m$. When the attack distance increases, $\theta_s$ becomes less flexible due to the increasing straight-line distance between the camera and the projector. The larger the distance is, the weaker the injected light becomes and the smaller the attack angle can be. On the other hand, the attack performance improves at night. We achieve the maximum 90\textdegree$ $ attack angle at $2-3 m$, which is larger than the vision sensors’ horizontal FOV (60\textdegree). 
Even when the light source is out-of-view, several orbs can still be produced, resulting in a successful orbs attack. 

The results indicate that: \emph{(1) the attack angle is more flexible at night than during the day due to the weak ambient light at night; (2) the range of attack angle becomes narrower with the larger attack distance in most of the cases due to the enlarged distance between the camera and the projector; (3) during the night, the orbs attack can forge fake depths even when the light source is out of the FOV.}
}

\begin{figure}[t]
\centering
	\includegraphics[width=0.45\textwidth]{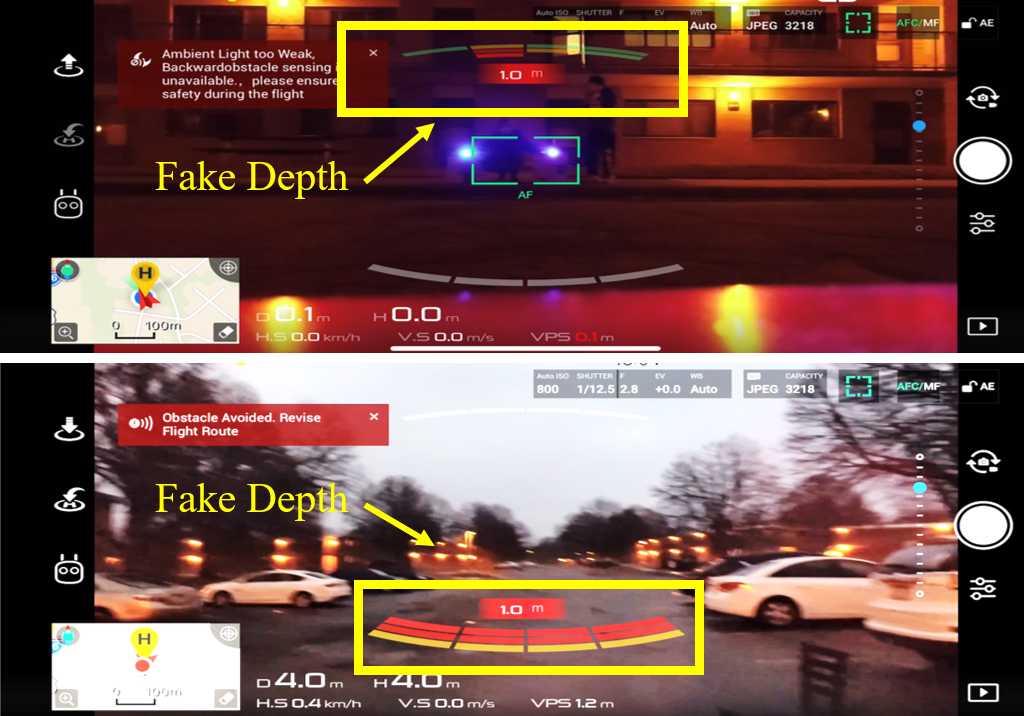}
	\caption{The attacks on forward and backward vision sensors on the DJI drone at night and in the day, respectively. The attack results are viewed on the controller.}
	\label{pic19}
\end{figure}

\rev{

\subsection{End-to-End Attack Validation}\label{endtoend}
For end-to-end attack validation, we first illustrate how the attacker can control the injected fake depth using the mathematical model. 
To showcase the real-world attack performance of \attack, we validate the end-to-end attack on both the flying DJI drone and RealSense camera. Please refer to Appendix~\ref{valiRealSense} for the RealSense attack validation. 

\textbf{Control and Validation of the Fake Depth.} An attacker can apply the mathematical model in Section \ref{attack_design} to control the fake depth generated at the victim device. For instance, with $d=1m$ and $b=0.12m$, the injected fake depth of X-shape beams attack at $4m$ away is $0.43m$ from Eq. (\ref{beam_x}).

In our experiments, the near-distance fake depths from X-shape beams attack and trapezoid-shape orbs attack are the ones we use to trigger the drone's OA, while the far-distance fake depths are created from trapezoid-shape beams attack and X-shape orbs attack. 
Since the step size of the depth in the drone's OA system is $0.5m$, we manually round the calculated values from mathematical model to its nearest step value.
Table~\ref{tab2} (second column) shows the expected near-distance fake depths from the mathematical model in comparison with the experimental results for X- and trapezoid-shape attacks. 
The results show that most of the injected fake depths from real experiments match with the expected ones. This indicates that the mathematical model can indeed be used to guide the attack process by adjusting the fake depths. 

\textbf{Sudden Braking.} To demonstrate the practicality of \attack, we launch the attack on a flying drone to induce sudden braking. We enable the P Mode on the DJI drone, with which it simply brakes and hovers when it detects the obstacle on its flying path. Fig. \ref{pic1} shows our attack towards a flying DJI drone from $7 m$ away. Fig. \ref{pic19} shows the attack effects on backward vision sensors during the daytime and forward vision sensors at night on the controller. We can see that the $1 m$ fake depth is detected in both cases. The drone starts sending warnings, and stops moving forward even though the pilot pushes the throttle forward on the controller. 

\textbf{Drifting Away and Shaking.} We then perform the experiments to drift the drone away from its original flying path by continuously injecting fake depths. 
DJI drone can automatically avoid obstacles rather than simple braking and hovering in some specific intelligent flying modes, \emph{i.e.,} ActiveTrack and TapFly~\cite{DJIUserManual}. We use ActiveTrack mode in our experiments, which allows the pilot to mark and track a moving object. To make the attack device more portable,
we use two high-lumen flashlights \cite{flashlight} to aim the drone. When the drone is tracking a subject at around $7m$ away from the attacker, we launch the attack and observe that the drone drifts away towards another direction. Moreover, we can slightly adjust the position of the light sources to change the fake depth locations from left to right alternately, which effectively shakes the drone. 
Please refer to our website for the attack demo. 

}

\rev{
\subsection{Attack Sensitivity Analysis}


\textbf{Aiming Precision.} The sensors mounted on flying drones and moving autonomous vehicles are usually tiny. Thus, aiming at the moving targets is quite challenging as shown in recent work~\cite{man2020ghostimage, cao2019adversarial, sun2020towards}, since the adversarial attack patterns should appear at the specific positions in the images or 3D point clouds. Unlike these efforts, our attack relaxes the requirements for precise aiming.

To aim the drone, an attacker has to track the target in real-time to ensure the light is projected into the appropriate sensors. The position of the beam is determined by the location of the projectors; the position of the orbs/glares in the 2D plane determines the position of the fake obstacle and its depth value in the 3D-depth map. 
To realize the attack in real attack scenarios, an attacker first visually estimates the distance between the two projectors and the stereo camera, and determines the distance between the two projectors based on the predetermined fake depth from the mathematical models. More importantly, our attack can be generalized on different devices by leveraging coarse-grained control of fake-depths using various attack patterns, e.g., X-shape beam attack generates near-distance fake-depth, whereas trapezoid-shape beam attack generates far-distance fake depth. In drones, a depth threshold is used to trigger OA, thus, a precise fake depth is not required. We experimentally validate that a coarse aiming precision is sufficient for a successful attack. 

Although the requirement of aiming precision is not high, the attacker does need to inject light into the camera. Particularly, in order to drift the drone away to follow a target trajectory, the attacker should closely follow the movement of the drone. Otherwise, when the light beams become out of the vision sensors' view, the attack could fail. 
On the other hand, with even the slightest movement of the lenses angle, a large difference can be observed on the attack target. Since the attacker can control the movement of the lenses, the attacker can aim the light beam at the moving target by slightly adjusting the angle of the lens. 
Based on the real-time feedback from the drone, e.g., its flying behavior or warnings, the attacker can adjust the attack angle to carry out a successful attack. 



\textbf{Other Factors.} 
\minor{We conduct numerous drone experiments under bad weather, such as snowing, raining, and windy conditions. The flying speed is $1~m/s$, and the flying height stays at 2-3 meters. The results show that our attack can work properly in all the experimented weather conditions, including a windy day with 20 miles per hour (mph) wind strength. However, the strong ambient light, fast speed, and high altitude of the flying drone could lead to attack failures. Specifically, when the ambient light is too strong  ($>4000 lux$), due to the auto exposure control, the vision sensors can clearly ``see" the background and the effect of the injected light beams or the lens flare effect weakens. Moreover, when the drone flies too fast, it becomes difficult to track the drone. Also, if the drone flies in a high altitude, a wider attack angle is expected as shown in Section~\ref{exp:position}, 
 leading to a lower attack success rate.

As a special case, when the expected fake depth falls into the focal length of the camera,
the attack will usually fail. However, since the glares/orbs cover multiple pixels, sometimes it will output the minimum depth value.

}
}

\section{Discussion}\label{discussion}
In this section, we discuss the practical challenges in launching \attack, and present the countermeasures. 

\subsection{Practical Considerations} \label{dis:generality}
\rev{
\textbf{Generality.} Stereo vision based depth estimation algorithms are widely adopted in OA systems. Nowadays, state-of-the-art depth estimation algorithms are AI-based~\cite{mayer2016large, pang2017cascade, chang2018pyramid, liang2018learning, yang2018segstereo, song2020edgestereo, yang2019drivingstereo}.
We test and show our attack against both traditional and AI-based stereo depth estimation algorithms in Section \ref{attack_depthAlg}. In fact, all the stereo depth estimation methods intend to accurately match stereo correspondence in the left and right images, which are susceptible to \attack. Besides, we confirm that the devices in our experiments use the most advanced AI-based depth estimation methods~\cite{depthCNN, ZED}. \attack can successfully attack these algorithms and generate fake depths, corroborating its generality. 

One limitation of our attack is the inability in adjusting pixels to precisely control fake-depth in physical attacks. However, as we mentioned earlier, we can earn a coarse-grained control of fake depths using different attack patterns.
\minor{Meanwhile, \attack attack can be generalized for different stereo cameras and drones, which only use  the stereo image pairs as the depth estimation input. However, if the autonomous systems, such as autonomous vehicles and certain types of drones, adopt other sensors (e.g., LiDAR, radar, acoustic sensors) for obstacle detection, \attack may not succeed as discussed in Section~\ref{subsec:counter}. In future, we will investigate the impact of \attack attack, when other types of sensors are used for obstacle detection.}
In summary, \attack is a black-box attack approach that does not rely on the knowledge of the target depth estimation algorithm or the victim device information. Therefore, \attack could cause a potential impact on a wide variety of OA systems. 


}
\rev{
\textbf{Conspicuousness.}
The appearance of glares in the stereo images seems to raise victims' awareness. 
However, the stereo images are not typically shown to the pilot/driver, such as DJI drones and Tesla cars \cite{Autopilot}. 
The highly concentrated beam also makes it harder for other observers to notice the attack, especially during daytime, e.g., as shown in Fig. \ref{pic9}, when the projector is on, the light beams are unnoticeable. At night, since the projection light becomes visible, a large attack angle may be useful to lower the vigilance of the victim. Additionally, the attack distance can reach up to $15m$ at night, which makes it difficult for humans to react and avoid attacks across such a long distance. Most importantly, the attacker can exploit the \emph{“safety first”} \cite{safetyfirst} rule in the autonomous systems
-- when it detects an obstacle, it will react to emergent situations preceding any human inputs to prevent life loss.
}



\textbf{Applicability.}
\attack is not only an attack approach, but it can also operate as a protection method against the privacy invasion threats posed by autonomous systems, such as drone flyovers \cite{PATRICK, TODAY, wbtv}. One particularly attractive application scenario is to use \attack to drift the drone away from a private property when a spying drone approaches the property. \attack can even shake the drone to avoid capturing clear videos or photos without damaging the drone, as opposed to the previous attempts using powerful lasers~\cite{lasernews} to damage the sensors, which may bring legal disputes.

\textbf{Limitation.} 
\rev{
One limitation of \attack is the portability of  attack equipment. To make our attack more flexible, the attacker can choose to use mini projectors or high-lumen flashlights which are more portable and light-weighted. However, the projection intensity of the mini projectors is usually low. As for the flashlight, due to its lower light frequency, 
flickering glares/orbs will be captured by cameras, which reduce 
the attack success rate. 
Essentially, there is a trade-off between the portability and attack success rate. Note that, laser can cause the damage of the camera image sensor, which is not suitable for launching a continuous attack. 
}

Another limitation is that \attack only works for one stereo camera at a time, but not for multiple devices simultaneously. To address this problem, we can add multiple pairs of light sources. However, it will be less convenient to launch the attack physically without multiple human operators. 

\subsection{Countermeasure}\label{subsec:counter}

\textbf{Using Sensor Fusion.}
\rev{
The most efficient countermeasure is to use sensor fusion \cite{wang2017sonic}, the art of combining multiple physical sensors to produce an accurate ``ground truth", even though each sensor might be unreliable on its own. Autonomous vehicles and drones are normally equipped with multiple cameras or other types of sensors, such as LiDARs and radars. In this work, we explore one type of sensor fusion to defend against \attack. We launch our attack on RealSense, which has both stereo vision sensors, infrared projectors \cite{ProjectorsD400}, and the advanced sensor fusion algorithm. 

In RealSense, the projected infrared light is patterned, the perceived pattern by the sensor can be used to extract the depth information. 
For example, if the pattern is a series of stripes projected onto a ball, the stripes perceived by the sensor would deform and bend around the surface of the ball in a specific way \cite{GuideToDepth}. Two types of data will become the input of the sensor fusion algorithm: one is from the structured light and coded light emitted by infrared projectors and captured by infrared sensors, and the other one is based on the images from stereo cameras. The algorithm fuses these two types of data and outputs a depth map.

\minor{In the experiment, we first disable the infrared sensors and launch our attack to inject a fake depth. Next, we enable the infrared sensors and sensor fusion. We repeat our experiments in both indoor and outdoor with various ambient light conditions and attack distances. The results show that the fake obstacle is present in all cases even under the active sensor fusion. This is likely caused by the strong light projection from the projector which may have washed out the projected infrared patterns, resulting in the failure of defense using sensor fusion.
This indicates that \attack has the potential to bypass 
sensor fusion algorithms.}

Sensor fusion can mitigate but not fully prevent all the safety issues, especially as the camera provides important inputs for depth estimation. Besides, since vehicle manufacturers endeavor to prevent the loss of lives, a \emph{“safety first”} approach is adopted for autonomous driving \cite{safetyfirst}, and as a result, the injected fake obstacles by \attack may still be treated as real objects even if some sensors in these sensor fusion algorithms disagree. 

\minor{Even though the results show that \attack can bypass the sensor fusion algorithms in RealSense, 
our attack could fail in cases when the stereo camera data is fused with other types of advanced sensors such as LiDAR  on autonomous vehicles, and we leave the further investigation as future work.}

\textbf{Detecting Over-Saturated Pixel.}
Under \attack, the injected glares/orbs are over saturated, so it could be a viable defense to detect over-saturated pixels in the stereo images. 
For example, Moizumi et al. propose to detect the traffic light by considering color saturation using in-vehicle stereo camera \cite{moizumi2016traffic}. This can be applied to defend against \attack. 
}

\minor{
 \textbf{Applying Film Polarizers.} Film polarizers  allow the light waves of a specific polarization to pass through and block light waves of other polarizations. 
 Applying a film polarizer on the camera lens might be a potential defense method. We experimentally verified that film polarizers can filter out the glares directly generated by the sunshine or other strong light sources \cite{FilmPolarizer}. However, they cannot cope with the lens flare effects or directly injected glares (light beams) presented in our attack.
}

\textbf{Adding Camera Lens Hood.}
Another complementary approach of defense is to add a camera lens hood \cite{Prevent_flare}, which can mitigate the lens flare effects to reduce the attack success rate. However, this is unfit for many autonomous systems, as the hood reduces the camera's FOV. 
Moreover, 
the lens hoods can only reduce the orbs generated by out-of-view light sources while they are unable to prevent the orbs created by in-view ones. Also, it cannot defend against the beams attack.

\textbf{Building Robust Neural Networks.}
Another line of defense would be to make neural networks themselves robust against computer vision-based adversarial attacks. However, the existing defenses proposed in \cite{papernot2016distillation, madry2017towards, lecuyer2019certified, wu2019defending} are ill-suited for our work, as \attack does not follow the constraints placed on traditional adversarial examples. Meanwhile, these defenses are not designed for arbitrarily large perturbations in a 3D environment. Moreover, even though the defense model proposed in \cite{nassi2020phantom} can defend against large perturbations in terms of 2D object classification, \attack could evade such defense as it targets at attacking the 3D depth perception. 
Designing a defense model towards the 3D adversarial attacks will be our future work.

\section{Related Work}\label{related_work}
In this section, we review related studies on sensor attacks towards drones.
With the rapidly growing popularity and capability of drones, sensor attacks towards drones have become a non-negligible security risk.
Son et al. are the first to investigate the sensor attack towards drones \cite{son2015rocking}. By injecting sound noise at the gyroscopes’ own resonant frequencies, the gyroscopes on the drones will fluctuate, leading to a DoS attack. Similarly, Wang et al. injected ultrasound signals around the gyroscope’s resonant frequencies to change the spinning speed of the four rotors to cause the DoS on the commercial drone \cite{wang2017sonic}. However, these two attacks can only achieve the non-continuous attack with the very short attack distance (around 10$cm$). 
Davidson et al. are the first to realize the continuous control of the drone by spoofing the optical flow sensor at the bottom of the drone. It can achieve the attack distance up to 3$m$ away in loiter mode, which is used to hold the drone precisely in the current position. Different from this work, 
\attack enables a diverse class of drone manipulations (i.e., stopping, shaking, drifting away), and achieves long-range continuous drone control under various flying modes.



\renewenvironment{comment}{}{
Sensor attacks towards cameras on autonomous driving vehicles have been investigated by researchers. Chen et al. are the first to propose and physically demonstrate different sensor attacks on autonomous vehicles \cite{yan2016can}. In their work, they use a laser to blind the camera and destroy the image sensors to cause the DoS attack. 
GhostImage \cite{man2020ghostimage} aims to attack the perception domain of the camera-based object detection and classification remotely. However, the requirement of precise aiming makes their attack impractical. Split-second attacks \cite{nassi2020phantom} utilize the weakness of the perception in computer vision algorithms to deceive the vision-based object detection and classification, thereby detecting the projected fake obstacles as real ones. 
Different from the previous work, we investigate a new attack genre on stereo cameras. 
}

\section{Conclusion}\label{conclusion}

In this paper, we present \attack, a new long-range attack towards the depth estimation-based OA systems on autonomous robotic vehicles and drones. By exploiting the vulnerabilities in the depth perception, an attacker can inject arbitrary lights into the stereo camera to create a fake obstacle depth. \attack consists of beams attack and orbs attack with three different attack patterns. We conduct the simulation using Ardupilot to demonstrate our attack towards drones.  
Through extensive real-world experiments, we find \attack is effective on different devices equipped with stereo cameras.
The successful long-range attacks against the flying DJI drone imply potential security impacts on different types of autonomous systems.

\section*{Acknowledgement}\label{sec:acknowledgement}

We thank the anonymous reviewers, our shepherd Dr. Sara Rampazzi for their valuable feedback on our work. We thank Yanmao Man for his helpful discussion on the attack setup. 
We also thank Chenhui Pan for his assistance during the experiments. This work is supported in part by the National Science Foundation grants CNS-1950171, CNS-1949753.


\bibliographystyle{plain}

\appendix
\section*{Appendix}
\section{Additional Experiments and Analysis}
\subsection{Green Orbs Visualization from RealSense}\label{green_orbs}

\begin{figure}[H]
\centering
	\includegraphics[width=0.42\textwidth]{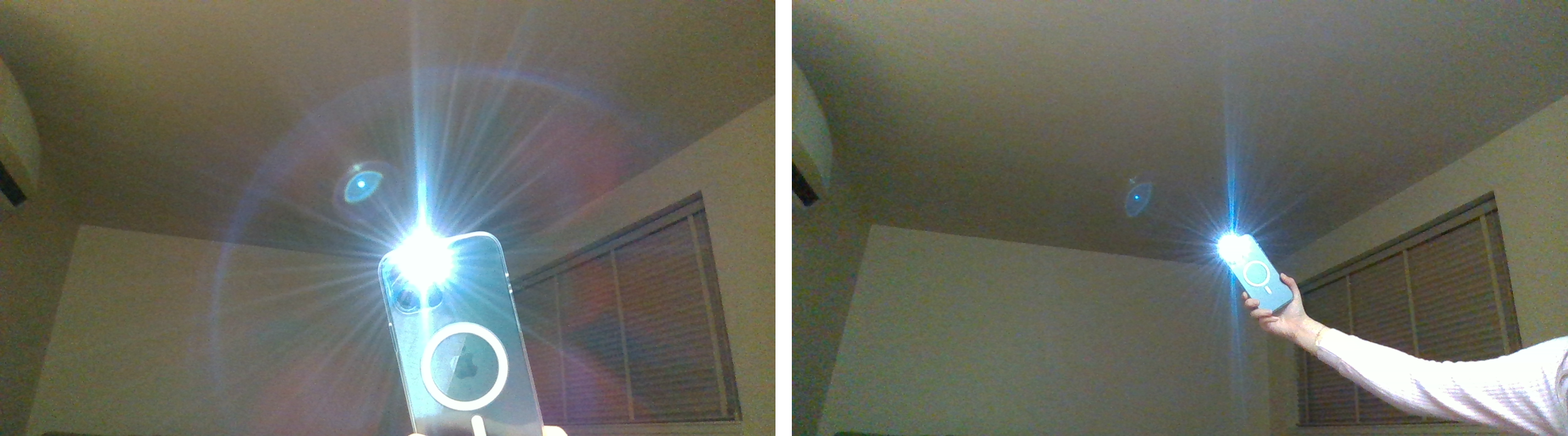}
\caption{Using an iPhone 12 Pro Max flash to project to RealSense with different distances and angles.} 
\label{pic20}
\end{figure}

As shown in Fig. \ref{pic20}, it can be seen that the green orb is always centrosymmetric to the light source. Moreover, the brightness of the orbs decreases as the projection distance increases.

\subsection{Beams Attack Unsuccessful Case}\label{unsuccess_case}

\begin{figure}[H]
\centering
	\includegraphics[width=0.31\textwidth]{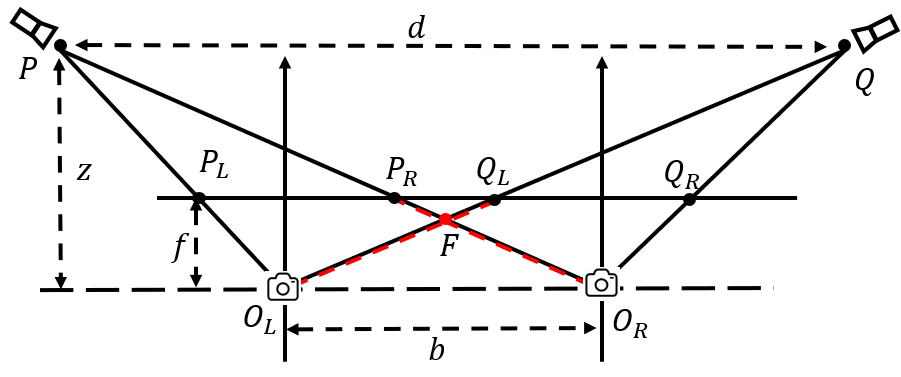}
	\caption{An unsuccessful case in X-shape beams attack when $b < d$.}
	\label{pic15}
\end{figure}

The analysis of the unsuccessful case of the X-shape beams attack is shown in Fig. \ref{pic15}. When $d \gg b$ or $z$ is too small, the fake depth is within the focal length of the camera, which is unrealistic in the optical imaging. Thus, the attack fails to generate the fake depth.

\subsection{Out-of-View Light Source}\label{out_of_view}

\begin{figure}[H]
\centering
\begin{minipage}{0.5\textwidth}
  \centering
\subfloat[Left and right images]{\includegraphics[width=0.465\textwidth]{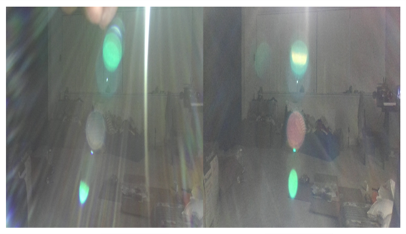}
\label{pic13a}}
\subfloat[3D point cloud]{\includegraphics[width=0.45\textwidth]{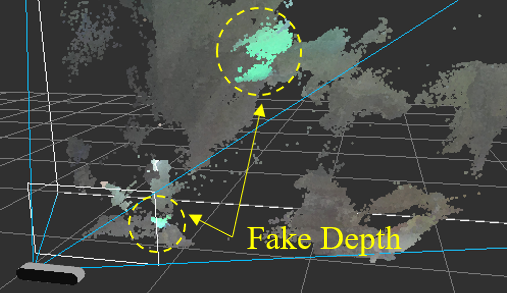}
\label{pic13b}}
\end{minipage}
\caption{The orbs attack on ZED. (a) shows the light sources are out of view, where several unclear orbs are generated; (b) shows several fake depths can be created consequently.} 
\label{pic13}
\end{figure}

In the experiments, we observe that out-of-view light sources can generate several undistinguished orbs, which can cause several fake depths on ZED as shown in Fig. \ref{pic13}.

\renewenvironment{comment}{}{
Contrasting with the trapezoid-shape orbs attack in Fig. \ref{pic4b}, we demonstrate the unsuccessful case with the same experimental setup as shown in Fig. \ref{pic16}. The result corroborates that beams attack cannot succeed when the distance between the projectors and the target distance is 1-2$m$ in our evaluation, while orbs attack is still successful.

\begin{figure}[H]
\centering
\begin{minipage}{0.5\textwidth}
  \centering
\subfloat[Left and right images]{\includegraphics[width=0.45\textwidth]{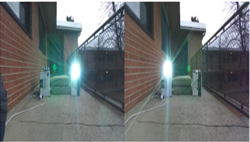}
\label{pic16a}}
\subfloat[Depth map]{\includegraphics[width=0.45\textwidth]{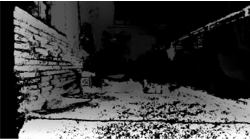}
\label{pic16b}}
\end{minipage}
\caption{The X-shape beams attack cannot create fake depth when the distance between projectors and stereo camera is 2$m$, in which case the expected fake depth is within the camera's focal length.} 
\label{pic16}
\end{figure}
}

\subsection{Validation on RealSense}\label{valiRealSense}
\begin{figure}[h]
\begin{minipage}{0.5\textwidth}
  \centering
\subfloat[RGB camera view]{\includegraphics[width=0.44\textwidth]{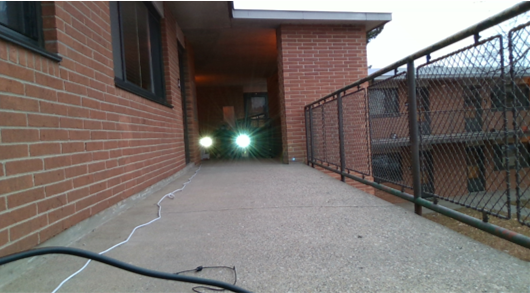}
\label{pic17a}}
\subfloat[Depth map]{\includegraphics[width=0.44\textwidth]{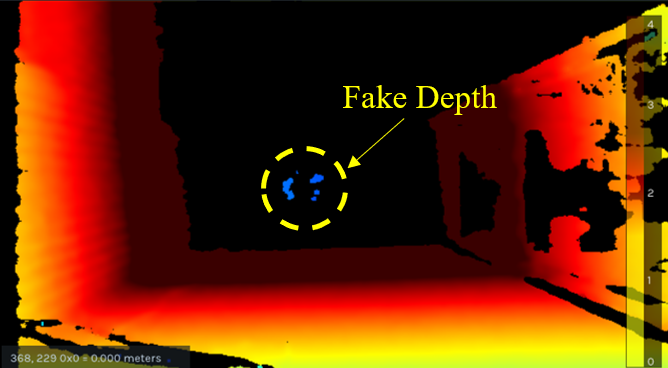}
\label{pic17b}}
\end{minipage}
\caption{(a) shows the view from the RGB camera under the attack. 
RealSense provides a colored depth map in (b) where the blue color indicates an object is at $0.5 m$.} 
\label{pic17}
\end{figure}

Fig. \ref{pic17} shows a near-distance fake depth (around 0.5$m$) can be generated when the attack distance is 10$m$ in RealSense. It demonstrates that \attack remains effective even without the access to its left and right stereo images. 

\rev{

\renewenvironment{comment}{}{

\begin{table*}[t]
\centering
\rev{
\caption{\rev{Expected fake depth from mathematical model.}}
\label{tab3}
\resizebox{0.75\textwidth}{!}{%
\begin{tabular}{|l|l|l|l|l|l|l|}
\hline
\multirow{2}{*}{\begin{tabular}[c]{@{}l@{}}\textbf{Attack} \\ \textbf{Distance ($m$)}\end{tabular}} &
  \multicolumn{2}{l|}{\begin{tabular}[c]{@{}l@{}}\textbf{Expected Fake Depth from} \\ \textbf{Beams Attack ($m$)}\end{tabular}} &
  \multicolumn{2}{l|}{\begin{tabular}[c]{@{}l@{}}\textbf{Expected Fake Depth from} \\ \textbf{Orbs Attack ($m$)}\end{tabular}} &
  \multicolumn{2}{l|}{\begin{tabular}[c]{@{}l@{}}\textbf{Realistic Expected Fake} \\ \textbf{Depth from Different Patterns ($m$)}\end{tabular}} \\ \cline{2-7} 
   & \textbf{X}    & \textbf{Trapezoid} & \textbf{X}     & \textbf{Trapezoid} & \textbf{X}      & \textbf{Trapezoid} \\ \hline
1  & 0.11 & -0.14     & -0.11 & 0.14      & 0.5    & 0.5       \\ \hline
2  & 0.21 & -0.27     & -0.21 & 0.27      & 0.5    & 0.5       \\ \hline
3  & 0.32 & -0.41     & -0.32 & 0.41      & 0.5    & 0.5       \\ \hline
4  & 0.43 & -0.55     & -0.43 & 0.55      & 0.5, 1 & 1         \\ \hline
5  & 0.54 & -0.68     & -0.54 & 0.68      & 0.5    & 1         \\ \hline
6  & 0.64 & -0.82     & -0.64 & 0.82      & 0.5    & 1         \\ \hline
7  & 0.75 & -0.95     & -0.75 & 0.95      & 0.5    & 1         \\ \hline
8  & 0.86 & -1.09     & -0.86 & 1.09      & 1      & 1         \\ \hline
9  & 0.96 & -1.23     & -0.96 & 1.23      & 1      & 1         \\ \hline
10 & 1.07 & -1.36     & -1.07 & 1.36      & 1      & 1.5       \\ \hline
11 & 1.18 & -1.5      & -1.18 & 1.5       & 1      & 1.5       \\ \hline
12 & 1.29 & -1.64     & -1.29 & 1.64      & 1.5    & 1.5       \\ \hline
13 & 1.39 & -1.77     & -1.39 & 1.77      & 1.5    & 2         \\ \hline
14 & 1.5  & -1.91     & -1.5  & 1.91      & 1.5    & 2         \\ \hline
15 & 1.61 & -2.05     & -1.61 & 2.05      & 1.5    & 2         \\ \hline
16 & 1.71 & -2.18     & -1.71 & 2.18      & 1.5    & 2         \\ \hline
\end{tabular}%
}
}
\end{table*}
}
}

\renewenvironment{comment}{}{
\subsection{Attack Setup at Night \minor{remove?}}

As shown in Fig. \ref{pic9}, two light beams are invisible for human during the daytime due to the concentration of the light beams. However, they are visible to human eyes at night as shown in Fig. \ref{pic22}.

\begin{figure}[t]
\centering
	\includegraphics[width=0.3\textwidth]{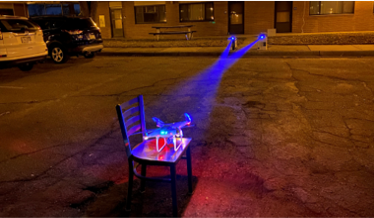}
	\caption{Experimental setup at night.}
	\label{pic22}
\end{figure}

\subsection{Practicality Demonstration \minor{remove?}}\label{appendix:demo}

As shown in Fig. \ref{pic23}, since the minor change of the lens angle can cause a significant light coverage difference on the victim side, the attacker can easily track the flying drone by continuously controlling the aiming angle of the taped lens. As long as the projection light is injected to the stereo camera, \attack can always succeed.

\begin{figure}[t]
\centering
\begin{minipage}{0.5\textwidth}
  \centering
\subfloat[Move the taped lens]{\includegraphics[width=0.65\textwidth]{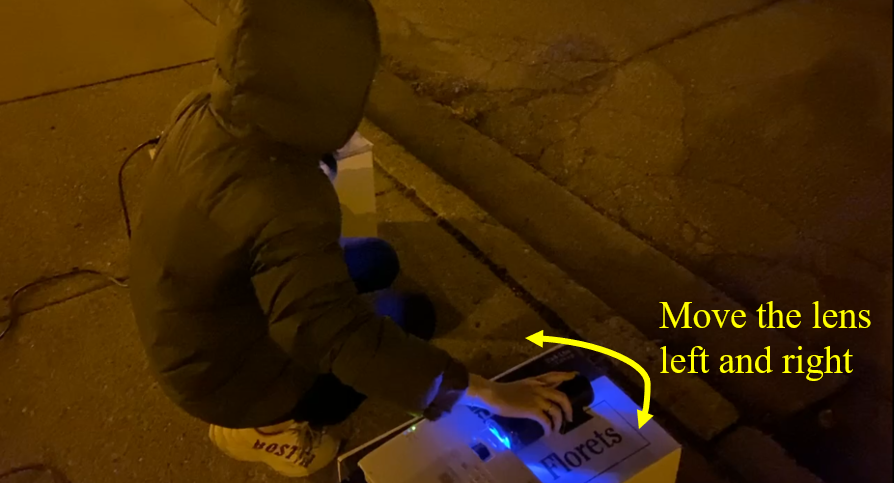}
\label{pic23a}}

\subfloat[Wide light coverage on DJI drone]{\includegraphics[width=0.67\textwidth]{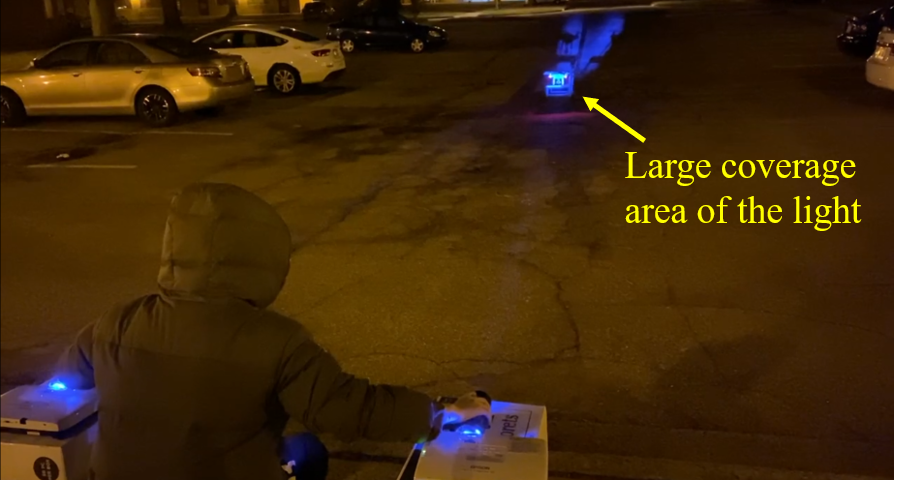}
\label{pic23b}}
\end{minipage}
\caption{The attacker slightly moves the taped lens left and right, and the wide light coverage on the DJI drone can be observed.} 
\label{pic23}
\end{figure}
}

\renewenvironment{comment}{}{
\rev{
\subsection{The Impact of Sensor Fusion}\label{impactSensorFusion}

Sensor fusion is the process of combining sensory data or data derived from disparate sources such that the resulting information has less uncertainty than would be possible when these sources were used individually \cite{SensorFusion}. To ensure correct and safe driving, sensor fusion is usually applied in autonomous driving system. In our experiments, we try to explore whether our attack is able to bypass sensor fusion. 

We launch our attack on RealSense, which has stereo vision sensors, infrared projectors \cite{ProjectorsD400}, together with the advanced sensor fusion algorithm (closed-source). In RealSense, the projected infrared light is patterned, the perceived pattern by the sensor can be used to extract the depth information. 
For example, if the pattern is a series of stripes projected onto a ball, the stripes perceived by the sensor would deform and bend around the surface of the ball in a specific way \cite{GuideToDepth}. Two types of data will become the input of the sensor fusion algorithm. One is from the structured light and coded light emitted by infrared projectors and captured by infrared sensors, and the other one is based on the images from stereo cameras. The algorithm fuses these two types of data and outputs a depth map.

\minor{In the experiments, we first disable the infrared sensors and launch our attack to inject a fake depth. Next, we enable the infrared sensors. We repeat our experiments in both indoor and outdoor with various ambient light conditions and attack distances. The result shows that the fake obstacle is still there without any change. The strong light from the projector might wash out the projected infrared pattern \cite{failurecase}, which might cause the failure of using sensor fusion to defend our attack.
This indicates that \attack has the potential to bypass 
sensor fusion algorithms. The evaluation of other types of sensor fusion algorithms is left as future work.}

Sensor fusion can mitigate but not fully prevent all the safety issues, especially as the camera provides important inputs for depth-estimation. Besides, since vehicle manufacturers endeavor to prevent the loss of lives, a \emph{“safety first”} approach is adopted for autonomous driving \cite{safetyfirst}, and as a result, the injected fake obstacles by \attack may still be treated as real objects even if some sensors in these sensor fusion algorithms disagree. 
}
}



\end{document}